\documentclass[journal]{IEEEtran}
\usepackage{graphicx}
\usepackage{subfigure}
\usepackage{epstopdf}
\graphicspath{{files/Images/}}
\DeclareGraphicsExtensions{.eps,.jpeg,.png}
\usepackage{color,soul}
\usepackage[cmex10]{amsmath}
\let\mybibitem\bibitem
\renewcommand{\bibitem}[1]{%
  \ifstrequal{#1}{nature}
    {\color{blue}\mybibitem{#1}}
    {\color{black}\mybibitem{#1}}%
}
\usepackage{xcolor,cite,etoolbox}
\makeatletter 
\pretocmd\@bibitem{\color{black}\csname keycolor#1\endcsname}{}{\fail}
\newcommand\citecolor[1]{\@namedef{keycolor#1}{\color{blue}}}
\makeatother

\usepackage{amsthm}
\usepackage{amsfonts}
\usepackage{cite}

\newtheorem{prop}{Proposition}
\newtheorem{obs}{Observation}

\usepackage[subnum]{cases}
\renewcommand{\qedsymbol}{\rule{0.7em}{0.7em}}
\newcommand{\h}{{$h$}}
\newcommand{\rr}{$R$}
\newcommand{\dr}{{$drone-BS$}}
\newcommand{\plmax}{$L_{max}$}
\newcommand{\arR}{$\mathcal{R}$}
\newcommand{\arW}{$\mathcal{W}$}
\newcommand{\arD}{$\mathcal{D}$}
\newcommand\numberthis{\addtocounter{equation}{1}\tag{\theequation}}
\newcommand{\sq}[2][0]{
  \mbox{$\medmuskip=#1mu\displaystyle#2$}%
}

\usepackage{algorithm}
\usepackage{algpseudocode}
\usepackage{pifont}

\newcommand{\mavi}[1]{{\leavevmode\color{blue} #1}}
\usepackage{tikz}

\newcommand\encircle[1]{%
  \tikz[baseline=(X.text)] 
    \node (X) [draw, shape=circle, inner sep=0, thick] {\strut #1};}

\DeclareRobustCommand\mytikzdot{\encircle}

\begin{document}
%
\title{Spatial Configuration of Agile Wireless Networks with Drone-BSs and User-in-the-loop}

\author{\IEEEauthorblockN{Irem Bor-Yaliniz,
Amr El-Keyi, and
Halim Yanikomeroglu}
\IEEEauthorblockA{}
\thanks{This work was supported in part by Huawei Canada Co., Ltd., in part by the Ontario Ministry of Economic Development and Innovations' ORF-RE (Ontario Research Fund Research
Excellence) program, and in part by the Natural Sciences and Engineering Council of Canada's (NSERC) Strategic Partnership Grants for Projects (SPG-P) program.} 
\thanks{The authors are with the Department of Systems and Computer Engineering,
Carleton University, Ottawa, Ontario, Canada (e-mail: \{irembor, amr.elkeyi,
halim\}@sce.carleton.ca).}}


\maketitle

\begin{abstract}
Agile networking can reduce over-engineering, costs, and energy waste. Towards that end, it is vital to exploit all degrees of freedom of wireless networks efficiently, so that service quality is not sacrificed. In order to reap the benefits of flexible networking, we propose a spatial network configuration scheme (SNC), which can result in efficient networking; both from the perspective of network capacity, and profitability. First, SNC utilizes the drone-base-stations (drone-BSs) to configure access points. Drone-BSs are shifting paradigms of heterogeneous wireless networks by providing radically flexible deployment opportunities. On the other hand, their limited endurance and potential high cost increase the importance of utilizing drone-BSs efficiently. Therefore, secondly, user mobility is exploited via user-in-the-loop (UIL), which aims at influencing users' mobility by offering incentives. The proposed uncoordinated SNC is a computationally efficient method, yet, it may be insufficient to exploit the synergy between drone-BSs and UIL. Hence, we propose joint SNC, which increases the performance gain along with the computational cost. Finally, semi-joint SNC combines benefits of joint SNC, with computational efficiency. Numerical results show that semi-joint SNC is two orders of magnitude times faster than joint SNC, and more than $15\%$ profit can be obtained compared to conventional systems.      
\end{abstract}

\begin{IEEEkeywords}
Unmanned aerial vehicle, user-in-the-loop, agile network, drone-BS.
\end{IEEEkeywords}

\IEEEpeerreviewmaketitle

\section{Introduction}
%
%
%
%
\IEEEPARstart{E}ver-increasing wireless demand is expected to grow in different dimensions due to extremely varying requirements of potential applications, from very low latency to very high data rate to very high energy efficiency. From tactile internet to internet of things, the future of wireless services is as exciting as challenging. Hence, densification of wireless networks seems to be inevitable~\cite{ultra_dense}. On the other hand, leaning to gross over-engineering comes at a high cost of not only CAPEX and OPEX\footnote{Capital expenses (CAPEX) and operational expenses (OPEX).}, but also environmental footprint. Therefore, flexible and agile wireless networking solutions, which can help reduce over-engineering without compromising quality, gained importance. Among these solutions, unmanned aerial vehicles, also known as drones, equipped with some functionalities of terrestrial base stations (\textit{drone-BSs}) recently attracted a significant amount of attention~\cite{bor_magazine}.

While drone-BSs add a degree of freedom to the deployment of wireless networks, another emerging research area is to exploit mobility of users. In the user-in-the-loop (UIL) paradigm, the network operator tries to influence the mobility of users by offering them incentives~\cite{uil_mag}. UIL schemes can be adjusted based on the target benefit of the network, be it energy efficiency or increasing revenue, which makes UIL another radically flexible networking technique.

In this paper, we investigate the synergy between the deployment of drone-BSs and designing the incentives for UIL schemes to improve efficiency and increase profitability of wireless networks. Cell association, traffic management, and load balancing can be thought of as related research areas. The difference between the previous schemes and this one is the following: Traditionally, the base station (BS) is fixed and the locations of users are random. In our case, the base station is mobile with a varying coverage area, and mobility of the users is \textit{influenceable} towards the benefit of the network. Hence, \textit{spatial network configuration (SNC)} can be performed.  
%
\subsection{Related Works}

Benefiting from agility of low-altitude drone-BS in providing on-demand capacity for wireless networks can make them the next frontier of heterogeneous wireless networks~\cite{bor_magazine}.  In most of the studies in the literature, which will be discussed shortly, either the altitude or the horizontal location of drone-BSs are assumed to be pre-determined, which makes the problem very similar to small-cell placement problems. In contrast, the approach in this paper is based on 3D placement of a drone-BS, i.e., jointly determining altitude and horizontal location of a drone-BS. This problem is introduced and efficiently solved in~\cite{bor2016efficient}. Then, a multi-objective 3D placement problem considering multi-tenancy, energy efficiency, caching and congestion release is formulated in~\cite{bor_magazine}, and caching is investigated futher in~\cite{chen2017caching}. Multi-tier drone-BS placement is investigated to show potential gains in spectral efficiency, throughput, latency and coverage in~\cite{sharma2016intelligent}, and a drone-BS network formation algorithm is developed in~\cite{park2016drone} drone by considering 3D placement. Spectrum sharing between single-tier and multi-tier drone-BSs is investigated to determine optimal density of drone-BSs in~\cite{7756327} by assuming pre-determined horizontal locations for drone-BSs. Terrestrial users are clustered to determine placement of drone-BSs in the horizontal domain in~\cite{d_elham}. In~\cite{lyu2016placement}, assuming a fixed coverage area for drone-BSs, a polynomial-time algorithm is developed to provide maximum coverage to a finite area with minimum number of drone-BSs. In~\cite{rupasinghe2016optimum}, optimum hovering positions of drone-BSs with antenna arrays are determined to minimize interference and maximize SNR. The proposed method is validated by exhaustive search, and provides computational efficiency and higher capacity performance by combining linear zero-force beamforming with transmit beamforming. Improving resilience of wireless networks is investigated in~\cite{6881196 ,7122576, hayajneh2016drone,narang2017cyber}. Downlink coverage analysis is conducted for a single drone-BS in~\cite{hayajneh2016drone}, and for a 3D drone-BS network in~\cite{chetlur2017downlink}. Although studies discussed so far consider hovering drone-BSs, an energy efficient trajectory is determined for point-to-point communications between a fixed-altitude drone-BS and terrestrial users in~\cite{d_zeng_2_2016energy}, and for drone-BSs with a fixed coverage area and altitude in~\cite{mozaffari2016mobile}. Other studies worth noting investigate the issues of releasing congestion, power allocation for drone-BSs, drone-BS placement via stochastic geometry-based network planning, association problem in C-RAN with drone-BSs, network performance analysis, drone-BSs as moving edge infrastructures, and development of placement and trajectory optimization algorithms from various aspects \cite{alzenad20173d, andryeyev2017increasing, faraj, yang2017proactive, shah2017association, becvar2017performance, zeng2017trajectory,lyu2017spectrum, mozaffari2017optimal, koyuncu2017deployment, rametta2017designing, mehtaaerial, wu2017joint}. Finally, in~\cite{galinina2016demystifying}, effects of mobile access points on business models are discussed. So far, only mobility of drone-BSs are exploited, however, in this study, we also exploit the mobility of the users by UIL.

\subsection{User Involvement in Network Operation}
User involvement is utilized as a method to improve system performance, with respect to user's perception or via user cooperation, for the purpose of application adaption, accurate crowd-sensing, improved cybersecurity and so on~\cite{8116483,bajaj_spectrum_2015,ozlu_user_2017,khalili_incentive_2015,pouryazdan_quantifying_2017,jaimes_survey_2015,haderer_preliminary_2013,giannetsos_trustworthy_2014,sheng_sensing_2013,ganti_mobile_2011,li_offloading_2017,su_incentive_2017,iosifidis_incentive_2014,saghezchi_incentive_2012,nunes_survey_2015,micro_1,micro_2,dumitrescu_design_2015,uil_mag,sawyer_u-beas:_2016,ometov_enabling_2016,andreev_network-assisted_2015,barbosa_architecture_2016,schoenen2017system,schoenen2018system,shan_user_2017,wang_novel_2018,suto_postdisaster_2017,du_data-driven_2018,wang_configurations_2018,mirahsan_admission_2018,mup2016,schoenen2011green,kasgari2018human,yang2012crowdsourcing,7996575,7306950,wang2017novel}. Nunes \emph{et al.} provide a survey in~\cite{nunes_survey_2015} where they argue that cyber-physical systems can benefit significantly from considering human element as a part of the system, instead of treating it as an uncontrollable external component. In fact, once the target benefits of networks are determined, main components of many systems with user involvement can be summarized as follows.
\begin{itemize}
\item Role of the user: The cooperative role of the user is the first building block of systems with user involvement. In general, the user may take actions, be passive, or systems with hybrid user roles are also possible~\cite{nunes_survey_2015}. 
\item Incentive design: According to the role of the user, incentives are designed. A good incentive design should be persuasive enough to convince users to cooperate with the network operator\footnote{The ``network operator'' is used here in a broader sense, such that it may be an actual network operator, e.g., AT\&T, Turkcell, Bell, or a person operating an application using the existing network, e.g., Foursquare, WhatsApp.}, and at the same time should maximize network's target benefits (e.g., green and profitable wireless network provisioning).  
\end{itemize}


Target benefit of the network determines the role and tasks of the users to be involved. Different roles can result in different system designs. Users can have a passive role, e.g., in many crowd-sensing applications such as~\cite{pouryazdan_quantifying_2017,haderer_preliminary_2013,sheng_sensing_2013,giannetsos_trustworthy_2014} users only provide their data. For instance, incentives are offered to users to improve crowd-sensing data quality in~\cite{pouryazdan_quantifying_2017}. In~\cite{kasgari2018human}, user perception is utilized to fine-tune QoS, which in turn provides energy savings. The study in~\cite{8116483} is an example of a hybrid system, where users can have both passive and active roles. For instance, in~\cite{8116483}, users either only watch or choose to share advertisements to obtain incentives. The target benefit in this study is to turn users into agents to distribute advertisement content, while developing sponsored data plans to create a win-win situation. Alternatively, users may be more actively involved by taking an action affecting the system performance, e.g., in case of UIL, the user may move to a position with better SINR~\cite{uil_mag}, or in~\cite{khalili_incentive_2015,ozlu_user_2017,bajaj_spectrum_2015} users share their resources with other users to support network operator's services.  In~\cite{bajaj_spectrum_2015}, licensed users trade under utilized spectrum resources with unlicensed users to improve spectrum utilization efficiency. In~\cite{ozlu_user_2017} and~\cite{khalili_incentive_2015}, users provide access points to improve connectivity of wireless networks. In the UIL method applied in~\cite{schoenen2011green,uil_mag,mirahsan2015user,schoenen2011user,sawyer_u-beas:_2016}, users receive monetary or non-monetary incentives (e.g., discount on service fees, improved service quality, reducing environmental impact of communications) so that the network operation can become more efficient, e.g., in terms of energy expenditure, wireless resource usage etc., and increase number of served users via \textit{demand shaping}~\cite{uil_mag}. For instance, users can receive incentives to delay their demand, or move to a better position where providing wireless services is more efficient. Hence, the users' demand can be shaped spatiotemporally. Note that the users can choose to comply with the offers or not, and making persuasive and profitable offers is key for the success of UIL systems. 

Incentive design has been puzzling researchers as a multi-disciplinary issue with complex factors. Many incentive methods have been developed, where they can be broadly categorized as follows~\cite{jaimes_survey_2015}.
\begin{itemize}
\item Non-monetary incentives: When the network operator provides incentives that are costless to the operator. These type of incentives heavily depend on the motivation of the user to volunteer. It is common to use games to increase motivation of users to participate~\cite{haderer_preliminary_2013}. 
\item Monetary incentives: When costly incentives are offered to the users, there is a trade-off between the persuasiveness of the offer and profitability of the operation. Moreover, even with monetary incentives, not all the users will comply~\cite{survey}. 
\end{itemize}

The studies on incentive methods either rely on hypothetical assumptions (e.g.,~\cite{pouryazdan_quantifying_2017,li_offloading_2017,wang_pricing_2018,8123043}), or field studies (e.g.,~\cite{survey,micro_1,micro_2,evers2014user}). Both methods suffer from reliability, as the statistical assumptions are hard to justify and many field studies are limited to a small group of participants with similar demographics, e.g., university students. Moreover, it is hard to generalize the results of a field study for all applications, since the user tasks and roles may differ. Nevertheless, in~\cite{micro_1} effectiveness of monetary examples are investigated based on an existing field study that has cumbersome requirements for the participants. This study with 36 participants suggests that if variable amounts of incentives are offered for similar tasks, the user persuasion rate to complete the task can be slightly higher than offering a uniform amount or a hidden amount that is revealed after completion~\cite{micro_1}. A similar study is conducted in~\cite{micro_2} with the purpose of crowd-sensing, i.e., users share their desired data in exchange with some amount of incentive. Similar to the study in~\cite{micro_1}, offering variable incentives resulted more and higher-quality data collection from a total of 55 participants. Note that the amount of incentives offered in these studies does not depend on the system requirements, rather they are static. On the other hand, studies  such as~\cite{yang2012crowdsourcing,7996575,7306950} consider dynamic factors to determine incentives, such as availability of resources, demand towards users' data, behavior of other users and so on. In~\cite{survey}, a unique survey is conducted with 100 participants (twice the amount of participants in previous studies) to understand the behavior of wireless network users in cooperating with the network operator. Certainly, the study in~\cite{survey} is inadequate to draw substantial conclusions on this complex issue, however, it is at least as comprehensive as the previous studies, and targeted towards operating wireless networks, rather than another field or purpose. Therefore, instead of using hypothetical assumptions about the user behavior, the results in~\cite{survey} is used for the analysis in this article.''\\ 
\subsection{Contributions and Organization}
Efficiently combining drone-BSs and UIL in cellular networks with the objective of maximizing profit while satisfying QoS requirements is a rather involved design problem. Both drone-BSs and UIL have their unique challenges~\cite{bor_magazine, uil_mag}. To the best of our knowledge, this paper presents the first system that considers both the mobility of the users and drone-BSs. The contributions of the paper can be summarized as follows:
\begin{itemize}
\item The gain from UIL for a uniform user distribution is analysed in Appendix~\ref{app:B}. Our analysis show that 50$\%$ relative gain can be obtained by incorporating UIL. Besides providing insights on the profitability of UIL, the analysis can be useful in cases of insufficient or incorrect information on user locations, and/or very high user mobility. 
\item When information on user location is available, the optimal incentive for an uncovered user at a given ground distance from the coverage of area of a base station is obtained in Section~\ref{sec:uil}.
\item A framework for SNC, consisting of 3D placement of a drone-BS, and incentive design for UIL, is presented. In Section~\ref{sec:USNC}, an \textit{uncoordinated}-SNC method is proposed to utilize the framework. It is shown that for a user, the gain from the UIL is strictly affected by user's proximity to the location of the drone-BS. 
\item In order to improve system performance by performing 3D placement with consideration to incentive design for UIL, a joint-SNC (JSNC) problem is formulated in Section~\ref{sec:JSNC}. The proposed JSNC problem is efficiently solvable via interior-point optimization, however, the computational complexity is very high.
\item A reduced-complexity joint-SNC problem, \textit{semi-JSNC}, is formulated in Section~\ref{sec:Semi}. Simulation results show that semi-JSNC can be more than 10 times faster than JSNC, while providing similar gain.  
\item All of the proposed SNC methods with varying computational complexities and accuracies, are shown to increase the profit of the network operator, as well as the number of served users by the drone-BS without increasing transmission power. 
\end{itemize}

The article is organized as follows. First, we describe the system model involving a drone-BS and a spatial UIL scheme in Section~\ref{sec:system_model}. Next, in Section~\ref{sec:USNC}, we discuss \textit{uncoordinated SNC} (USNC), where first the drone-BS is positioned, and then the UIL incentives are designed. In Section~\ref{sec:JSNC}, we discuss \textit{joint SNC} (JSNC), where the placement and incentive design is determined simultaneously. Since the resulting problem has high computational complexity, we introduce a semi-JSNC method in Section~\ref{sec:Semi}. Finally, we present simulations and results in Section~\ref{sec:sim}, and conclude the paper in Section~\ref{sec:conc}.

\section{System Model}\label{sec:system_model}
We consider a scenario in which the existing infrastructure of mobile network operators is temporarily insufficient to respond to the demand in a finite region, \arW, containing a set of users, $\mathbb{U}$. Insufficiency of the network may be the result of overloading, malfunction, or a similar unexpected situation. We assume that the location of each user in 2-D horizontal space, $(x_i, y_i) \ \forall i\in\mathbb{U}$, and quality of service (QoS) requirement of the users in terms of maximum tolerable path loss, $\gamma$, are known. A drone-BS is to be utilized to support the network as shown in Fig.~\ref{fig:generic} by offloading as many users as possible from the network, while satisfying the QoS requirements and maintaining profitability of the network. In particular, the aim is to position the drone-BS in a 3D location such that the number of covered users and the benefit from UIL can be maximized. While the drone-BS is positioned, only users whose QoS requirements cannot be satisfied by the terrestrial-BS, i.e., unserved users, are considered. Since these users would not be served otherwise, they handover to the drone-BS, if they are in the coverage region of drone-BS. Main factors contributing to handover are better RSSI (due to line-of-sight and dynamic positioning with respect to users' locations) and bandwidth availability (due to orthogonality of resources of the drone-BS)~\cite{chetlur2017downlink,bor2016efficient,alzenad20173d,sboui2017achievable,7122576,rupasinghe2016optimum,8368062,8292783,park2016drone,8269064,chen2017caching,lyu2016placement,sharma2016uav}.
\begin{figure}
\includegraphics[width=0.5\textwidth]{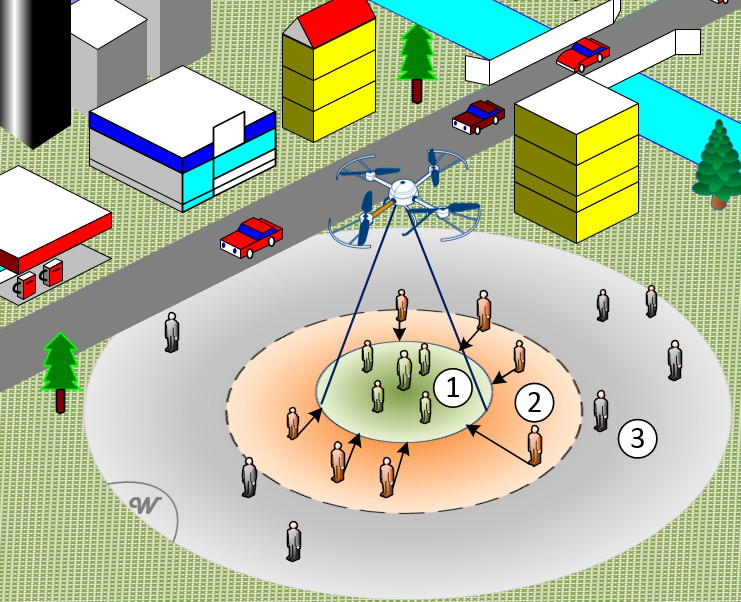}
\caption{Description of SNC: Users in region \mytikzdot{1} are covered without additional cost. Users in region \mytikzdot{2} are within the extended coverage region, therefore, they receive an incentive offer to move towards the coverage area. Users in region \mytikzdot{3} cannot be covered even with UIL, and hence do not receive an offer.}
\label{fig:generic}
\end{figure}
%
Assume the network incorporates a SNC scheme, which suggests certain displacements to each user. Incentives are offered to the users in the extended coverage area for persuading them to accept suggested movements. In the proposed SNC scheme, suggested displacement is towards the coverage area of a drone-BS, where users receive better service compared to their present location. Since the inter-cell distance is large in rural areas, which makes UIL of the SNC scheme less useful, we consider urban environments. Note that, in this network, both the source (drone-BS), and the sinks (users), are mobile. Therefore, \textit{spatial network configuration} (SNC) can be performed by determining the following set of parameters,
\begin{enumerate}
\item optimal location of the drone-BS in 3-D space, $(x_D, y_D, h_D)$,
\item suggested displacement of each user $i$, $\Delta_i = [\Delta x_i, \Delta y_i]$, and
\item incentives offered to each user $i$, $\tau_i$.
\end{enumerate}
The aim is to determine these values such that the efficiency and profitability of the network is maximized. Assuming that the drone-BS has orthogonal resources to the existing infrastructure, details of the channel and UIL models are discussed in the following sections.
\subsection{Air-to-Ground Channel Model}
Air-to-ground channel models differ from terrestrial channel models, because the altitude of a drone-BS affects the amount of path loss. The effect of altitude is reflected in the probability of having a line-of-sight (LOS) link between a user and a drone-BS. This probability depends on the altitude of the drone-BS, the horizontal distance between a user and the drone-BS, and the type of the deployment environment (rural, urban etc.)~\cite{al-hourani_optimal_2014}. The studies considering probability of LOS links for air-to-ground channels are still at their infancy, however, in a widely used model\footnote{The model is used in the following publications and many others: \cite{7756327, hayajneh2016drone, mozaffari2016mobile, mozaffari_drone_2015, 7390451, chen2016caching}.}, the probability of having LOS link is given by~\cite{al-hourani_optimal_2014}
\begin{equation}
P(h_D,r_i) = \frac{1}{1+a\exp\left(-b\left(\frac{180}{\pi}\tan^{\!-\!1}\left(\frac{h_D}{r_i}\right)-a\right)\right)},
\label{eq:plos}
\end{equation}%
where $r_i =  \sqrt{(x_D - x_i)^2 + (y_D - y_i)^2 }$ is the horizontal distance between the drone-BS at $(x_D, y_D, h_D)$ and the $i$th user located at $(x_i, y_i)$, and $a$ and $b$ are parameters of the environment. Environmental parameters depend on the average characteristics of urban areas, such as the density and average height of the buildings~\cite{al-hourani_optimal_2014}. Then, using~\eqref{eq:plos}, the path loss between the drone-BS and the $i^{th}$ user is calculated as 
\begin{equation}
L(h_D, r_i) = 20\log\left(\sqrt{h_D^2 + r_i^2 }\right) + z_1P(h_D,r_i)+ z_2,
\label{eq:pl2}
\end{equation} 
where $z_1$ and $z_2$ are constants such that $z_1 = \eta_{\text{LoS}} - \eta_{\text{NLoS}}$, and $z_2 = 20\log(\frac{4\pi f_c}{c})+ \eta_{\text{NLoS}}$, $f_c$ is the carrier frequency (Hz), $c$ is the speed of light (m/s), $\eta_{LoS}$ and $\eta_{NLoS}$ (in dB) are respectively the losses corresponding to the LoS and non-LoS connections depending on the environment. Parameters of four different environments are provided in Table~\ref{tab:param}. Note that, \eqref{eq:plos} and \eqref{eq:pl2} state that the path loss of air-to-ground channels depends on the the location of the drone-BS in 3D space. 
\begin{table}
\caption{Channel model parameters of different environments}
\label{tab:param}
\centering
\begin{tabular}{|c|c|}
\hline
\textbf{Environment} & \textbf{Parameters ($a$, $b$, $\eta_{\text{LoS}}$, $\eta_{\text{NLoS}})$}\\
\hline
Suburban & (4.88, 0.43, 0.1, 21)\\
\hline
Urban & (9.61, 0.16, 1, 20)\\
\hline
Dense Urban & (12.08, 0.11, 1.6, 23)\\
\hline
\end{tabular}
\end{table}
\subsection{User-in-the-loop Model}
\label{sec:uil}
UIL is a closed loop model with the user as the system to influence for increasing the efficiency of wireless networks. UIL is based on offering incentives to users to persuade them to take specific actions, such as delaying their demand in time, or changing their location in space. These incentives can be discounts on the service price, higher data rates, reducing carbon footprint (green networking), or even penalties for refusing the change~\cite{uil_mag}. In that sense, UIL is a hybrid user involvement method, where both monetary and non-monetary offers can be provided to users. The UIL method employed in this article assumes a spatial UIL scheme with monetary incentives. As a result, although at a cost, potential wastes of energy and capacity can be prevented, resulting in more efficient networks.

UIL can be used either spatially, where the user moves in space to obtain better links with the destination (higher spectral efficiency), or temporarily, in which case the user delays the demand in time. Assuming a drone-BS is opportunistically utilized for a specific time period, and the objective is to offload as many users as possible to the drone-BS during its utilization, a spatial UIL scheme is considered in this study. 

In order to investigate the reaction of the users to various UIL schemes for data, video and voice services, a detailed survey is conducted in~\cite{survey}. For data services, the probability of a user $i$ accepting the incentive, $\tau_i$, which is a discount on the cost of the service for user $i$, and moving a distance less than or equal to $d_i = \sqrt{\Delta x_i^2 + \Delta y_i^2}$ is given as 
\begin{equation}
\text{P}\lbrace \text{user $i$ moves distance }\leq \, d_i\rbrace = \mathrm{e}^{-\beta(\tau_i) d_i},
\label{eq:prob_move}
\end{equation}
where  $\beta (\tau_i)$ is the \textit{persuasion parameter}. 

\begin{figure*}
  \centering
  \subfigure[]{\includegraphics[width=0.4\textwidth]{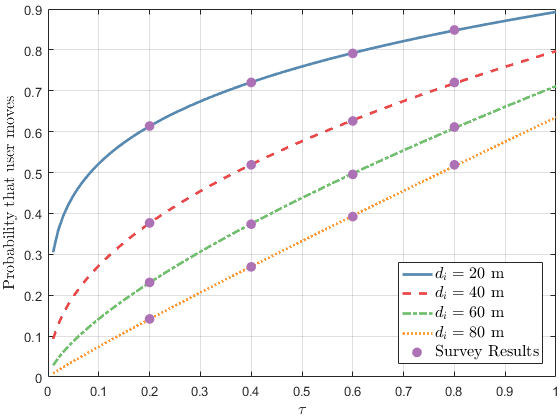}\label{fig:incentive}}
  \subfigure[]{\includegraphics[width=0.5\textwidth]{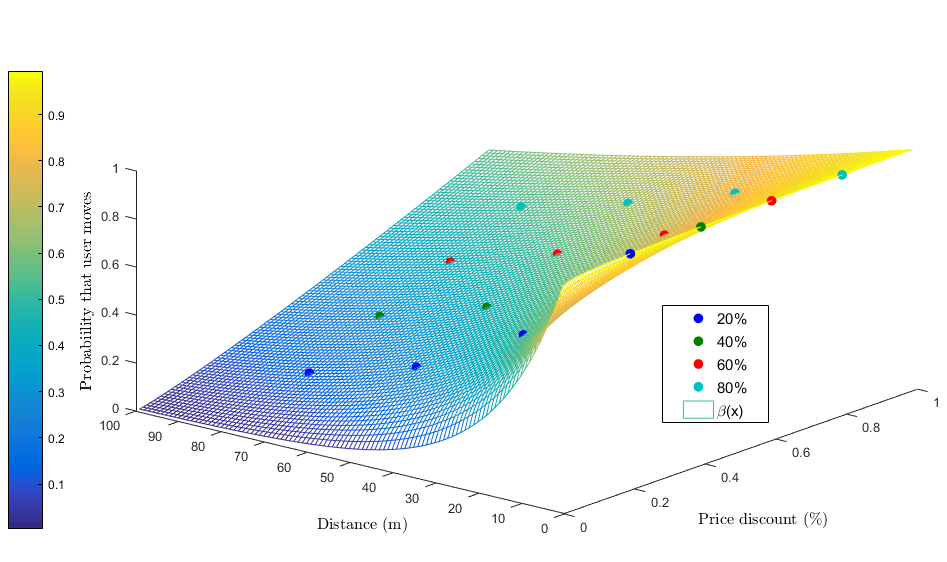}\label{fig:beta3}}
 \caption{(a) Probability of a user accepting to move with respect to incentive (percentage of price discount) and distance: Survey data is fit to $\beta(\tau)$ in~\eqref{eq:beta}. (b) 3D representation of survey results and $\beta(\tau)$ regarding a user's probability to move with respect to price discount and distance.}
  \label{fig:generic}
\end{figure*}

In~\cite{survey}, $\beta(\tau)$ is calculated based on four different incentives, namely, $ \tau\!=\!0.2,\, 0.4,\, 0.6,$ and $0.8$, corresponding to discount amounts of $\!20\%,\, 40\%,\, 60\%,$ and $80\%$, respectively. These incentive amounts  provide the persuasion values, $\beta(\tau)\!=0.0244,\, 0.0164,\, 0.0117,$ and $0.0082$, respectively. A continuous and tractable function for $\beta(\tau)$ is obtained by logarithmic curve fitting in our study, such that
\begin{equation}
\label{eq:beta}
\beta(\tau) = k_1\ln(\tau)+k_2,
\end{equation} 
where $k_1 = -0.01166$ and $k_2 = 0.005676$ for $\tau \in [0,1]$, where $0$ and $1$ corresponds to ``no incentive" and ``free service" ($100\%$ off), respectively. The root mean square error (RMSE) of~\eqref{eq:beta} is $8.359\times10^{-5}$ for the logarithmic fitting function with only 2 parameters. Since an RMSE value that is closer to zero indicates a good fit, this value shows that~\eqref{eq:beta} is highly likely to provide an accurate prediction.
In Fig.~\ref{fig:incentive}, the probability of being persuaded in~\eqref{eq:prob_move} is shown both with the points from the survey in~\cite{survey}, and the fitting in~\eqref{eq:beta}. Also, Fig.~\ref{fig:beta3} shows the continuity and fit of the approximation from a wider perspective. It is observed that \eqref{eq:beta} is a proper fit for the provided values.  

Proper selection of UIL system parameters, such as $\tau_i$ and $d_i$, has a key role in the performance of any UIL scheme~\cite{survey, mirahsan2015user, uil_mag, sawyer_u-beas:_2016, uil_ziyang}. The \textit{incentivized} user movement suggestion may be a function of a variety of elements, including user type, application type and urgency, the need for UIL, and so on. It may provide less ossified extension regions, on the contrary to Fig.~\ref{fig:generic}. For instance, if there is a compliant user e.g., a student type user, the user may be given an incentive even when located in region 3. On the other hand, a non-compliant user, e.g., a business type user, may not receive incentives even if the user is located in region 2. Note that, compliance level of a user may change if the application is highly valuable to the user at that moment, e.g., a video conference call, or the the application will be running for a while. By utilizing big data (history) and machine learning technologies, the incentivized move suggestions can be made user/context/situation-aware~\cite{schoenen2017system}. However, there is not an existing tractable model to incorporate all these aspects.

As discussed earlier in Section~\ref{sec:system_model}, the SNC parameters not only include UIL, but also involves dynamic placement of a drone-BS. In this case, the model in~\cite{survey} provides simplicity and tractability. SNC parameters can either be determined separately by considering drone-BS and UIL systems in an uncoordinated fashion, or they can be determined jointly. 
%
%
%
\section{Uncoordinated Spatial Network Configuration}\label{sec:USNC}
A system incorporating drone-BSs and UIL requires handling two processes for spatial network coordination:
\begin{itemize}
\item 3D placement of a drone-BS, i.e., determining $(x_D, y_D, h_D)$, and
\item Incentive design for user-in-the-loop, i.e., determining $\tau_i$ and $d_i$ for each user $i \in \mathbb{U}$.
\end{itemize}
The method of handling these processes in a sequential manner is termed \textit{uncoordinated spatial network coordination} (USNC). In the following sections, both steps will be discussed in detail. 

\subsection{3D Placement of a Drone-BS}
\label{sec:3d}
Placement of a drone-BS is different from terrestrial BS placement because of the following reasons:
\begin{enumerate}
\item In addition to choosing the drone-BS's location in the horizontal space $(x_D, y_D)$, we need to determine its altitude, $h_D$, as well.
\item The coverage area of a terrestrial BS is known a priori. However, the coverage area of a drone-BS depends on its altitude, and is unknown before solving the placement problem.
\item The mobility of the drone-BS allows it to move wherever the demand is, rather than terrestrial cells waiting for the demand to come towards them. As a result, the coverage region providing the maximum benefit to the network should be found.
\end{enumerate}
The first item indicates that the placement of the drone-BS is a 3D problem. In addition, the last two items, which are determining the size of the coverage area, and identifying the location of the coverage area must be considered jointly. 
We assume that a user is in the coverage region of the drone-BS if the air-to-ground link satisfies its QoS requirement. Hence, user $i$ is served by the drone-BS, if $L(h_D, r_i) \leq \gamma$. Using~\eqref{eq:pl2}, we can re-write this condition as
\begin{equation}
h_D^2 + r_i^2 \leq 10^\frac{\gamma - (z_1P(h_D,r_i) + z_2)}{10}.
\label{eq:new_gamma}
\end{equation}
Let $u_i \in \{0,1\}$ denote a binary variable that indicates whether user $i$ is served by the drone-BS, or not. Using the variable $u_i$, which is equal to 1, only if the user $i$ is served by the drone-BS, and equal to 0 otherwise, the following constraint,
\begin{equation}
u_i(h_D^2 + r_i^2) \leq 10^\frac{\gamma - (z_1P(h_D,r_i) + z_2)}{10},
\label{eq:new_gamma}
\end{equation}
determines whether user $i$ is covered, or not. This constraint can be further manipulated to
\begin{equation}
\begin{aligned}
\sqrt{h_D^2 + r_i^2}\leq \sqrt{10^\frac{\gamma - (z_1P(h_D,r_i) + z_2)}{10}} + M_1(1-u_i),
\label{eq:either}
\end{aligned}
\end{equation}
where $M_1$ is a constant that is slightly larger than the maximum possible value of the distance between a user and the drone-BS. Observe that when $u_{i} = 1$, \eqref{eq:either} is equivalent to \eqref{eq:new_gamma}. If $u_i = 0$, since $M_1$ is large enough, this constraint is released. Now, we can continue by determining the objective function.

Assuming a fixed QoS for all users, the best region to be served by the drone-BS is identified with the maximum number of users covered. By using~\eqref{eq:either}, the placement problem can be written as
\begin{align*}\label{eq:opt}
&\underset{x_D,y_D,h_D,\{u_i\}}{\text{maximize}}
\qquad \sum_{i \in\mathbb{U}}u_i\nonumber\\
& \text{s.t.} \
\sqrt{h_D^2 + r_i^2}\leq \sqrt{10^\frac{\gamma - (z_1P(h_D,r_i) + z_2)}{10}} + M_1(1-u_i),\nonumber
\\ 
&\qquad \qquad \qquad \qquad \qquad \qquad \qquad \qquad \ \ \forall i = 1,...,|\mathbb{U}|,\nonumber\\
& \ \ \quad x_{l} \leq x_D \leq x_{u}, \\
& \ \ \quad y_{l} \leq y_D \leq y_{u}, \numberthis \\
& \ \ \quad h_{l} \leq h_D \leq h_{u},\nonumber\\
& \ \ \quad u_{i} \in \{0,1\},
\qquad \qquad \qquad \qquad \qquad \forall i = 1,...,|\mathbb{U}|,
\end{align*}
where $|\cdot|$ represents the cardinality of a set, subscripts $(\cdot)_l$ and $(\cdot)_u$ denote respectively the minimum and maximum allowed values for $x_D$, $y_D$, and $h_D$ of the drone-BS. Note that there are quadratic, exponential and binary terms in this problem, which makes it a mixed integer non-linear problem (MINLP). %
%
We will show that this problem can be solved efficiently by using a combination of the interior-point optimizer of MOSEK solver and bisection search.

Let $R$ denote the radius of the area to be covered by the drone-BS. Then, if the user $i$ is covered, $r_{i} \leq R$ must be satisfied, i.e., the served user must be located within the coverage region. Let $\alpha$ denote the ratio of the altitude to the coverage region, such that
\begin{equation}
\alpha = \frac{h_D}{R}.
\label{eq:alpha}
\end{equation}
Then, \eqref{eq:either} can be re-organized by using~\eqref{eq:alpha} 
\begin{equation}
r_i \leq \Gamma(\alpha) - M_2(1-u_i),\label{eq:r_leq_gamma}
\end{equation}
where $M_2$ is a constant similar to $M_1$, and
\begin{equation}
\Gamma(\alpha) = \sqrt{\frac{10^\frac{\gamma - (z_1P(\alpha) + z_2)}{10}}{(1+\alpha^2)}},
\label{eq:Gamma}
\end{equation}
where
\begin{equation}
P(\alpha) = \frac{1}{1+a\exp\left(-b\left(\arctan\left(\alpha\right)-a\right)\right)}
\end{equation}
by~\eqref{eq:plos}. 

\begin{obs}\label{obs:1}
For any QoS requirement, $\gamma$, and for any operating frequency, $f_c$, if a local maxima, $\Gamma^*$, exists in the function $\Gamma(\alpha)$ defined in~\eqref{eq:Gamma}, then it is the only local maxima of the function for $\alpha \in (0, \infty )$ for the propagation environments whose parameters are listed in Table~\ref{tab:param}. The $\alpha^*$ yielding $\Gamma^*$ is the solution of
\begin{equation}
\sq{\alpha\pi \sq{\!\left(\!a\mathrm{e}^{\!-\!b\!\left(\!\frac{\!180\!\tan^{\!-\!1}\left(\!\alpha\!\right)}{{\!\pi}}\!-a\!\right)}+1\!\right)^2}}\!\!-\!\sq{k_3}\mathrm{e}^{\!-\!b\!\left(\frac{\!180\!\tan^{\!-\!1}\left(\!\alpha\!\right)}{{\pi}}\!-\!a\!\right)} = 0,\label{eq:alpha_star}
\end{equation}
where $k_3 = -9\ln\left(10\right)abz_1$.
\end{obs}

For an explanation, please refer to Appendix~\ref{app:A}.\\

The largest feasible set for Problem~\ref{eq:opt} can be obtained by substituting $\Gamma(\alpha)$ with $\Gamma^*$, which is the only maxima of $\Gamma(\alpha)$. Since no closed-form expression is available by~\eqref{eq:alpha_star}, bisection search can be used to obtain a numerical value for $\alpha^*$. Then, substituting $\Gamma(\alpha)$ with $\Gamma^*$ yields the following mixed integer quadratically constrained problem (MIQCP)
\begin{align*}\label{eq:opt_final}
&\underset{x_D,y_D,\{u_i\}}{\text{maximize}}
\qquad \sum_{i \in\mathbb{U}}u_i\nonumber\\
& \text{s. t.} \ r_i \leq \Gamma^* + (1-u_i)M_2,
\qquad \qquad \quad \, \forall i = 1,...,|\mathbb{U}|,\nonumber\\
& \ \ \quad x_{l} \leq x_D \leq x_{u}, \\
& \ \ \quad y_{l} \leq y_D \leq y_{u}, \numberthis \\
& \ \ \quad u_{i} \in \{0,1\},
\qquad \qquad \qquad \qquad \quad \ \  \forall i = 1,...,|\mathbb{U}|.
\end{align*}
While $R$ can be derived from the location of the covered user with maximum $r_i$, $h_D$ can be calculated by~\eqref{eq:alpha}. Hence, $h_D$ is omitted from~\eqref{eq:opt_final} as a variable. Therefore, both complexity of calculation, and the number of variables are reduced. Then, this problem can be solved efficiently by using interior-point optimizer of MOSEK. 
%
%
The above problem formulation yields the optimal position of the drone-BS.
Once the drone-BS is positioned, $d_i$ can be calculated for each non-covered user, and incentives can be designed to increase revenue. 

\subsection{Incentive Design for User-in-the-loop}\label{sec:single_user}
Based on the UIL scheme in Section~\ref{sec:uil}, the average \textit{unit profit} obtained from a user at a distance $d_i$ meters from the coverage region after receiving incentive $\tau_i$ is 
\begin{equation}
\Pi(\tau_i, d_i) =  (1-\tau_i)\mathrm{e}^{-\beta(\tau_i)d_i}.
\label{eq:profit}
\end{equation}
\begin{prop}
For an uncovered user at a ground distance $d_i$ from the coverage area of the drone-BS, the optimal incentive that maximizes $\Pi(\tau_i, d_i)$, $\tau_i^*$, is given by 
\begin{equation}
\tau_i^* = \frac{k_1d_i}{k_1d_i-1}.
\label{eq:tau_star}
\end{equation}
\end{prop}
\begin{proof}
For a given $d_i$, the stationary point of $\Pi(\tau_i, d_i)$, $\tau_i^*$, can be evaluated by taking the first derivative of~\eqref{eq:profit} and equating it to zero, which yields~\eqref{eq:tau_star}.
The second derivative of~\eqref{eq:profit} is given by
\begin{equation}
\!\resizebox{0.90\columnwidth}{!}{$\displaystyle{\ \mathrm{e}^{\!-\!k_2 d_i}k_1 d_i \tau_i^{-\!(k_1 d_i + 1)}\!\left(\! (k_1 d_i+1) \tau_i^{\!-\!(k_1 d_i + 1)}\! -\! (k_1d_i\! -\!1)\!\right).}$}
\label{eq:second_derivative}
\end{equation}
It can be shown that~\eqref{eq:second_derivative} is negative for all values of $d_i$ at $\tau_i^*$. Hence, $\Pi(\tau_i, d_i)$ is nondecreasing for $\tau_i \leq \tau_i^*$, and nonincreasing for $\tau_i \geq \tau_i^*$ for all $\tau_i \in (0,1]$, which satisfies the second-order condition for quasiconcavity of $\Pi(\tau_i, d_i)$ in $\tau_i$ for all values of $d_i$~\cite{boyd}. Therefore, $\tau_i^*$ is the global maximum point of~\eqref{eq:profit}. 
\end{proof}

\begin{figure}
\centering
\includegraphics[width=0.48\textwidth]{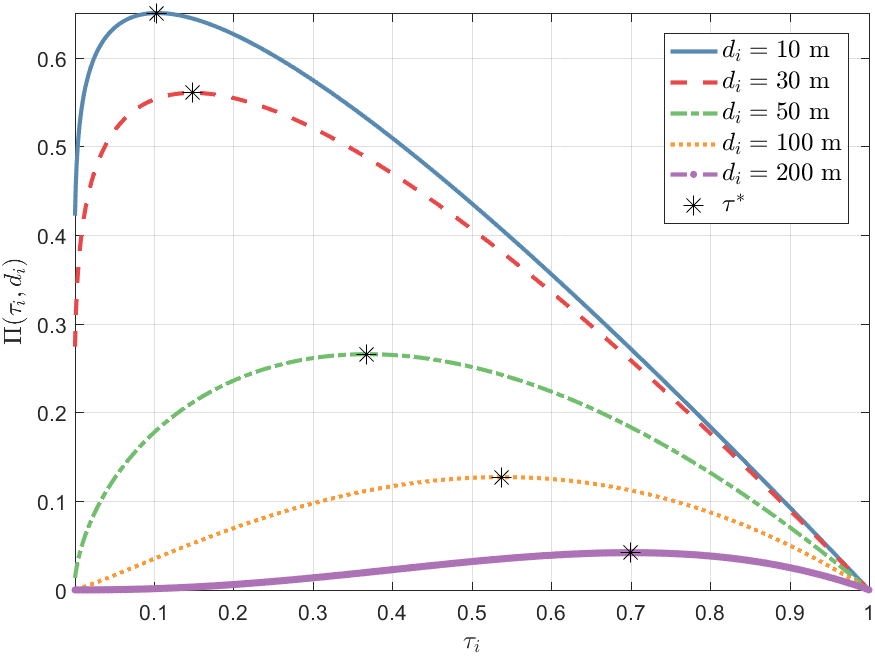}
\caption{Unit profit vs. incentive for several distances.} 
\label{fig:unit_profit}
\end{figure}

Once the drone-BS is placed as explained in Section~\ref{sec:3d}, $d_i$ can be calculated, and the above analysis can be used to offer the optimal incentive to each user. In Fig.~\ref{fig:unit_profit}, $\Pi(\tau_i, d_i)$ is shown for users at several distances. It can be observed that the users in proximity can become profitable with small incentive offers, however, the users that are far from the drone-BS do not provide much gain. For instance, $10.4\%$ discount offer can provide $65\%$ average profit for a user that should move $10$m. On the other hand, offering $53.8\%$ discount can only provide $12.7\%$ average profit for a user that should move $100$m, as the likelihood of accepting the incentive is very low. Therefore, it can be critical to position the drone-BS by considering the persuasion parameter and profitability from UIL, which is discussed next.

\section{Joint Spatial Network Configuration}\label{sec:JSNC}
Differently from the previous section, a \textit{joint spatial network configuration} (JSNC) problem is proposed to simultaneously position the drone-BS in 3D and determine the incentive to be offered to each user. Hence, mobility of both the drone-BS and users can be jointly considered as degrees of freedom of the spatial configuration of mobile networks. 

\begin{figure}[!h]
\includegraphics[width=0.45\textwidth]{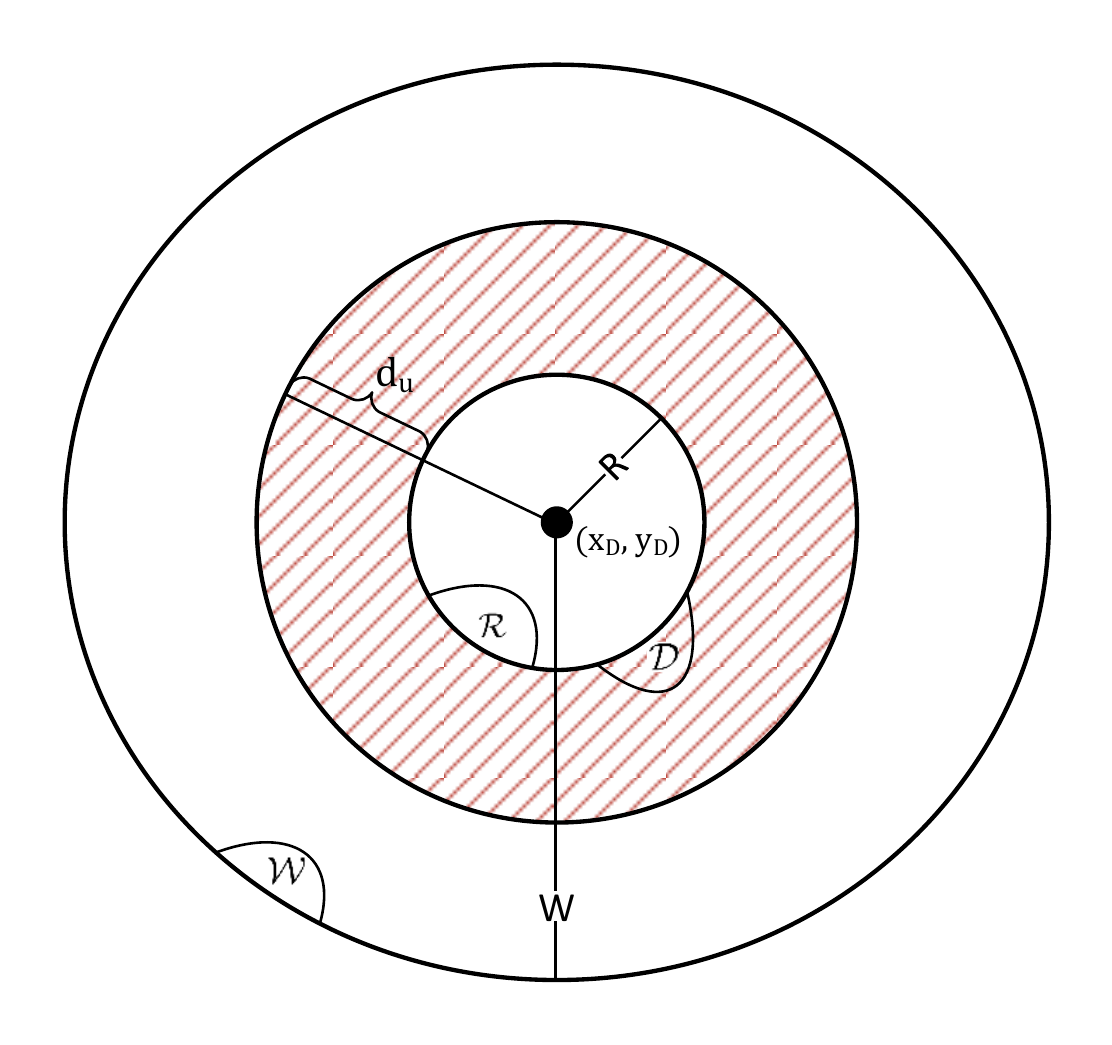}
\caption{Coverage area of a drone-BS ($\mathcal{R}$), area where the UIL model is utilized ($\mathcal{D}$), and a finite space where users are distributed ($\mathcal{W}$).}
\label{fig:range}
\end{figure}

\label{sec:problem}
In order to make UIL an inherent part of the placement problem, the coverage condition in the first constraint of~\eqref{eq:opt_final} can be modified as 
\begin{equation}
r_i-d_i \leq \Gamma^*\!+\!u_i(1\!-\!M_3),
\label{eq:cov2}
\end{equation} 
where $M_3$ is a slightly larger value than the maximum possible value of the right-hand-side (RHS) of \eqref{eq:cov2}. 
In contrast to the approach in Section~\ref{sec:3d}, maximizing the number of covered users does not necessarily mean maximizing profit this time, because additional users come at a cost with UIL. In fact, the coverage condition in~\eqref{eq:cov2} yields three regions in a circular area, \arW,~as shown in Fig.~\ref{fig:range}:
\begin{enumerate}
\item The circular coverage area of the drone-BS at $(x_D, y_D)$ with radius $R$ is defined as the region $\mathcal{R}$. 
\item The shaded area, $\mathcal{D}$, is where the UIL model is utilized. It is obtained by excluding $\mathcal{R}$ from a concentric circular area with the radius of $R+d_u$, where $d_u$ is an upper bound on the distance of a user from the coverage area of the drone-BS such that the users further than $d_u$ meters do not receive incentive offers. 
\item \arW-$(\mathcal{R} \cup \mathcal{D})$ is the region that cannot be served by the drone-BS even with UIL. 
\end{enumerate}
Note that, \eqref{eq:pl2} yields a circular coverage area for a drone-BS, as well as, designing incentives based on $d_i$ yields a circular expansion area by UIL (in Section~\ref{sec:uil}). Hence, a homogeneous environment is maintained. Further analysis on coverage of a drone-BS in a system with UIL and \textit{regional incentive design} for uniformly distributed users can be found in Appendix~\ref{app:B}. 

The profit obtained from a user $i$ based on the coverage region can be written as
\begin{numcases}{\Upsilon(\tau_i,d_i):= \label{eq:piece}}
1, \ \qquad \qquad \quad \ \ \ \text{ if $(x_i,\, y_i)\in \mathcal{R}$}, \label{eq:inR}\\
\Pi(\tau_i, d_i), \qquad \quad \ \text{ if $(x_i,\, y_i) \in \mathcal{D}$, }\label{eq:inD}\\
0, \ \ \  \qquad \qquad \qquad \quad \quad \text{ o.w.}\label{eq:out} 
\end{numcases}
where \eqref{eq:inR}, \eqref{eq:inD} and \eqref{eq:out} indicate the users covered in \arR, coverage in \arD, and outage, respectively. Hence, a JSNC problem for maximizing profit can be formulated as %
%
\begin{subequations}\label{eq:fal}
\begin{align}
&\underset{\parbox{5.5em}{\centering \scriptsize $x_D$, $y_D$, $\{u_i, d_i, \tau_i\}$}}{\text{maximize}}
\quad \sum_{i \in \mathbb{U}}\Upsilon(\tau_i,d_i)\tag{\ref{eq:fal}}\label{o:fal}\\
& \text{s. t.}
\ r_i-d_i \leq \Gamma^*\!+\!u_i(1\!-\!M_3),\label{c:fal1} \\
& \qquad \qquad \qquad \qquad \qquad \qquad \qquad  \quad \ \forall i = 1,...,|\mathbb{U}|,\nonumber\\
& \qquad  u_id_i \leq d_{u}, \qquad \qquad \qquad \qquad \ \ \forall i = 1,...,|\mathbb{U}|,\label{c:fal3}\\
& \qquad x_{l} \leq x_D \leq x_{u},  \label{c:fal4}\numberthis\\
& \qquad y_{l} \leq y_D \leq y_{u}, \label{c:fal5}\\
& \qquad  u_i \in \{0,1\},
\qquad \qquad \qquad \qquad \  \forall i = 1,...,|\mathbb{U}| \label{c:fal7},
\end{align}
\end{subequations}
where $d_u$ represents the upper bound for the amount of displacement.
%
Similarly to~\eqref{eq:opt}, $M_3$ in~\eqref{c:fal1} releases the condition of coverage, if the user is too far to be covered. If a user is persuaded to move towards the coverage region of drone-BS, \eqref{c:fal1} ensures that the user satisfies the condition of coverage in~\eqref{eq:cov2}. Note that, the actual displacement of the user is $\sqrt{(x_i-\Delta x_i)^2 + (y_i-\Delta y_i)^2 }$.
However, instead of $\Delta x_i$ and $\Delta y_i$, $d_i$ indicating the amount of movement is used to decrease number of variables in the problem. Once the problem is solved, $\Delta x_i$ and $\Delta y_i$ can obtained from $(x_i, y_i)$ and $(x_D, y_D)$. Unfortunately, the problem in~\eqref{eq:fal} is not efficiently solvable, due to \eqref{eq:inD}, which is a non-linear multivariate function, and requires further manipulations. 
\begin{figure}[!t]
\centering
\includegraphics[width=0.48\textwidth]{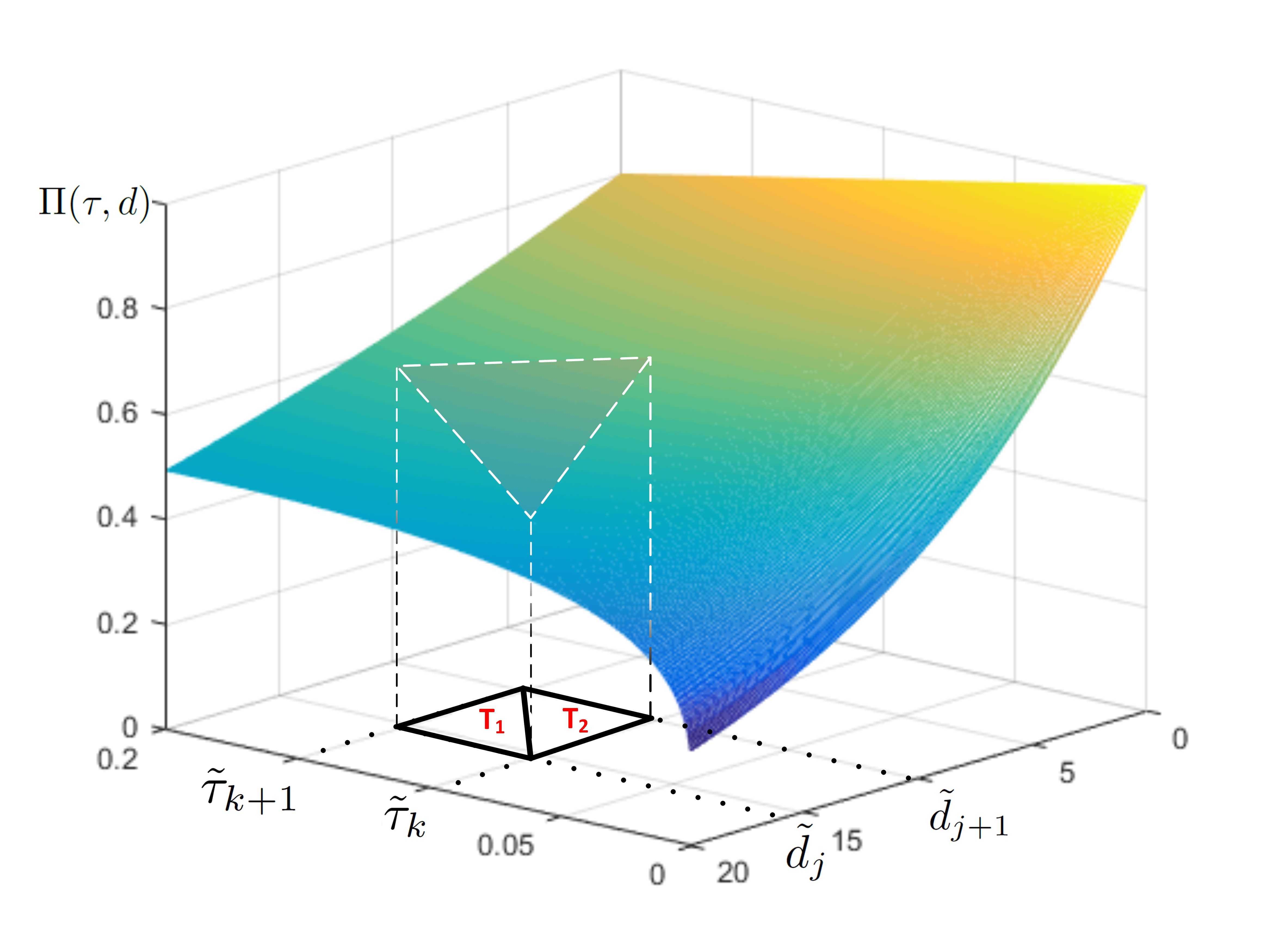}
\caption{Illustration of the triangle method on a portion of $\Pi(\tau,d)$.} 
\label{fig:trig}
\end{figure}
The \textit{triangle method} is one of the most popular methods for piecewise linear approximations of multivariate functions~\cite{dambrosio_piecewise_2010}. For $\Pi(\tau_i, d_i)$, the triangle method partitions the $\tau$ and $d$ axes into $m$ and $n$ sampling intervals, respectively. Let $\{\tilde{\tau_k}\}_{k=1}^m$ and $\{\tilde{d_j}\}_{j=1}^n$ represent incentive and displacement vertices, respectively. In order to approximate a given $\Pi({\tau}_i, {d}_i)$, the rectangular area with vertices satisfying $\tilde{\tau}_k\leq {\tau}_i \leq \tilde{\tau}_{k+1}$, and $\tilde{d}_j \leq {d}_i\leq \tilde{d}_{j+1}$ is found for $k = 1,...,m$ and $j = 1,..., n$. As shown in Fig.~\ref{fig:trig}, the rectangular area can be divided into two triangles, $T_1$ and $T_2$, with the vertices given by $\{(\tilde{\tau}_k,\tilde{d}_j), (\tilde{\tau}_k, \tilde{d}_{j+1}), (\tilde{\tau}_{k+1}, \tilde{d}_{j+1})\} $ and $\{(\tilde{\tau}_k,\tilde{d}_j), (\tilde{\tau}_{k+1}, \tilde{d}_{j}), (\tilde{\tau}_{k+1}, \tilde{d}_{j+1})\} $. Only one of the triangles contain $({\tau}_i, {d}_i)$. Therefore, the $T_1$ in Fig.~\ref{fig:trig} is chosen if the following is true,
\begin{equation}
{d}_i > \tilde{d}_j + ({\tau}_i - \tilde{\tau}_k)\frac{\tilde{d}_{j+1} - \tilde{d}_j}{\tilde{\tau}_{k+1} - \tilde{\tau}_k},
\label{eq:triangle}
\end{equation}
and $T_2$ is chosen otherwise. 

In general, the point to be approximated for each user $i$ can be anywhere in the search space, and the approximation points are unknown. Therefore, all the points on the grid must be considered. A mixed integer linear program (MILP) can be formulated to approximate $\Pi(\tau_i, d_i)$ :
\begin{subequations}\label{eq:search}
\begin{align}
& \sum\limits_{k=1}^n \sum\limits_{j=1}^m \rho_{i,k,j}\tilde{\tau}_k = \tau_i, \qquad \, \qquad \qquad \forall i = 1,...,|\mathbb{U}|,\label{c:rho2}\\
& \sum\limits_{k=1}^n \sum\limits_{j=1}^m \rho_{i,k,j}\tilde{d}_j = d_i, \qquad \, \qquad \qquad \forall i = 1,...,|\mathbb{U}|,\label{c:rho3}\\
& \sum\limits_{k=1}^n \sum\limits_{j=1}^m \rho_{i,k,j}\Pi(\tilde{\tau}\!_k, \tilde{d}_j\!) = \Pi(\tau_i, d_i), \quad  \forall i = 1,...,|\mathbb{U}|,\label{c:rho4}
\end{align}
\end{subequations}
where $\rho_{i,k,j}$ indicate continuous variables that are introduced for each sampling point. In order to choose three indices corresponding to a triangle, $\rho_{i,k,j}$ must be a special ordered set of type 3 (SOS3), which means that only 3 consecutive elements of the set can be non-zero~\cite{kallrath2012algebraic}. Also for a user $i$, 
\begin{equation}
\sum\limits_{k=1}^n \sum\limits_{j=1}^m \rho_{i,k,j} = 1,
\label{eq:lamunu}
\end{equation}
and each $\rho_{i,k,j} \in [0,1]$.
\begin{figure}[!t]
\includegraphics[width=0.48\textwidth]{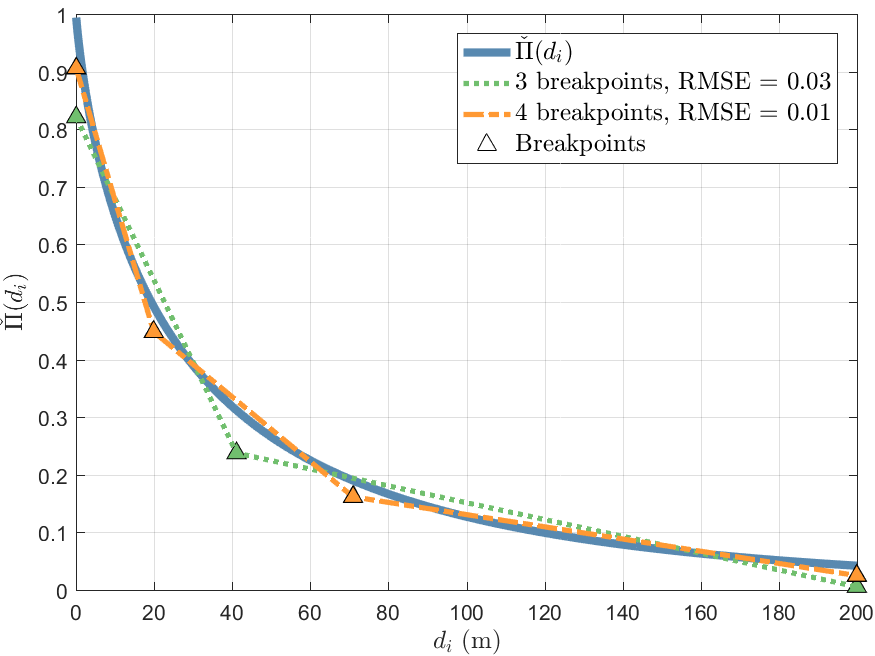}
\caption{Piece-wise approximation with 3 and 4 breakpoints, and the resulting RMSE values, where $d_u = 200$m.}
\label{fig:rmse}
\end{figure}

For instance, considering Fig.~\ref{fig:trig}, $\Pi(\tau_i, d_i)$ can be approximated by the following convex combination of the values of $\Pi(\tau, d)$ at the vertices of the chosen triangle,
\begin{equation}
\rho_{i,k,j} \Pi(\tilde{\tau}_k, \tilde{d}_j) + \rho_{i,k\!+\!1,j\!+\!1}\! \Pi(\tilde{\tau}_{k\!+\!1}, \tilde{d}_{j\!+\!1}) +\! \rho_{i,\hat{k},\hat{j}} \Pi(\hat{\tau}, \hat{d}),
\label{eq:cvx_rho}
\end{equation}
where the vertices of the diagonal, $(\tilde{\tau}_k,\tilde{d}_j)$ and $(\tilde{\tau}_{k+1}, \tilde{d}_{j+1})$ are common in both triangles, $(\hat{\tau}, \hat{d})$ indicate the third set of vertices differentiating the $T_1$ and $T_2$ based on~\eqref{eq:triangle}, and $\rho_{i,\hat{k},\hat{j}}$ indicate the weights corresponding to these vertices.
Formulating the optimization problem for the triangle method does not suffice to approximate the objective function of~\eqref{eq:fal} because it only represents~\eqref{eq:inD}. However~\eqref{eq:inR} and~\eqref{eq:out} too must be represented in the problem formulation. For that reason, in addition to $u_i$ in~\eqref{eq:fal}, consider another binary variable, $w_i$, indicating whether the user is in \arD~or not. These two variables, $u_i$ and $w_i$ can be used to represent all regions in~Fig.\ref{fig:range}, such that
\begin{itemize}
\item $u_i = 1$ indicates that the user is in \arR,
\item $w_i = 1$ indicates that the user is in \arD,
\item $u_i + w_i = 0$ indicates that the user is not covered,
\item $u_i+w_i \leq 1$ ensures that the user is either in \arR~or in \arD.
\end{itemize}
Finally, using triangle method, we can formulate~\eqref{eq:fal} as the following optimization problem:
\begin{subequations}\label{p:upsi}
\begin{align}
&\underset{\parbox{5.5em}{\centering \scriptsize $x_D,$ $y_D,$ $\{u_i, w_i, d_i, \tau_i, \Pi_i^a\}$, $\{\rho_{i,k,j}, \kappa^1_{i,k,j}, \kappa^2_{i,k,j}\}$ }}{\text{maximize}}
\qquad \sum_{i \in \mathbb{U}}(u_i + \Pi^a_i)\tag{\ref{p:upsi}}\label{ob:upsi}\\
& \text{s. t.}
\ \ \sq{r_i \leq \Gamma^* + M(1-u_i),}
\quad \qquad \quad \ \ \ \forall i = 1,...,|\mathbb{U}|,\label{c:coverage}\\
&\qquad \, \sq{r_i-d_i  \leq \Gamma^* + M(1-w_i),}
\quad \quad \ \ \ \ \forall i = 1,...,|\mathbb{U}|,\label{c:coverage2}\\
& \qquad  \sum\limits_{k=1}^n \sum\limits_{j=1}^m \rho_{i,k,j} = w_i,  \quad\qquad \qquad \, \ \forall i = 1,...,|\mathbb{U}|,\label{c:rho1}\\  
& \qquad  \sum\limits_{k=1}^n \sum\limits_{j=1}^m \rho_{i,k,j}\tilde{\tau}_k = \tau_i, \ \ \qquad \qquad \forall i = 1,...,|\mathbb{U}|,\label{c:rho2}\\
& \qquad  \sum\limits_{k=1}^n \sum\limits_{j=1}^m \rho_{i,k,j}\tilde{d}_j = d_i,\ \ \qquad \qquad \forall i = 1,...,|\mathbb{U}|,\label{c:rho3}\\
&\qquad \sum\limits_{k=1}^n \sum\limits_{j=1}^m \rho_{i,k,j}\Pi(\tilde{\tau}\!_k, \tilde{d}_j\!) = \Pi^a_i, \quad \ \  \forall i = 1,...,|\mathbb{U}|,\label{c:rho4}  \\
& \qquad  \sum\limits_{k=1}^{n-1} \sum\limits_{j=1}^{m-1} \kappa^1_{i,k,j} + \kappa^2_{i,k,j} = 1, \qquad \ \forall i = 1,...,|\mathbb{U}|,\label{c:kap1}\\
& \qquad  \rho_{i,k,j} \leq \kappa^1_{i,k,j} +\kappa^2_{i,k,j} + \kappa^1_{i,k,j-1} + \kappa^2_{i,k-1,j}\label{c:kap2}\\
&\qquad  + \kappa^1_{i,k-1,j-1} +\kappa^2_{i,k-1,j-1}, \qquad \ \ \forall i = 1,...,|\mathbb{U}|,\nonumber\\
&\qquad  \qquad  \qquad  \qquad \quad \ \, (k = 1,...,n), \ \ (j = 1,...,m),\nonumber\\
& \qquad  u_i + w_i \leq 1,\qquad \qquad \qquad \qquad \ \, \forall i = 1,...,|\mathbb{U}|,\label{c:inside}\\
& \qquad  u_i, w_i \in \{0,1\},\qquad \qquad \qquad \quad  \ \, \forall i = 1,...,|\mathbb{U}|,\\
& \qquad  \rho_{i,k,j} \in [0,1], \qquad \qquad \qquad \quad \ \ \ \,  \forall i = 1,...,|\mathbb{U}|,\\
&\qquad \qquad \qquad \qquad \quad \ (k = 1,...,n), \ \ \,(j = 1,...,m), \nonumber\\
& \qquad  \kappa^1_{i,k,j}, \kappa^2_{i,k,j} \in \{0,1\}, \qquad \qquad \quad \, \forall i = 1,...,|\mathbb{U}|,\\
&\qquad \qquad \qquad \qquad \quad \ \, (k = 1,...,n), \ \ (j = 1,...,m), \nonumber\\
& \qquad  x_{l} \leq x_D \leq x_{u}, \numberthis\\
& \qquad  y_{l} \leq y_D \leq y_{u},
\end{align}
\end{subequations}
where $\rho_{i,k,j}$ corresponds to the weights sampling points~\eqref{eq:cvx_rho}, $\kappa^1_{i,k,j}$ and $\kappa^2_{i,k,j}$, which are 0 at the extremes, allow choosing either $T_1$ or $T_2$, respectively. Utilizing $\kappa^1_{i,k,j}$ and $\kappa^2_{i,k,j}$ in constraints~\eqref{c:kap1} and~\eqref{c:kap2} provides $\rho_{i,k,j}$ variables to form a SOS3 type set. Hence, three vertices forming a triangle can be chosen. Although full formulation is provided here for completeness, $\kappa^1_{i,k,j}$ and $\kappa^2_{i,k,j}$ are not needed, if a solver allowing special ordered sets is used. Note that, a constraint similar to~\eqref{c:fal3} is not required, because the vertices already limit possible values of $d_i$.

Constraints~\eqref{c:rho1} through \eqref{c:rho4} are used to approximate $\Pi_i^a$ using the triangle method. Note that, \eqref{c:rho1} corresponds to~\eqref{eq:lamunu}, where the RHS of~\eqref{eq:lamunu} was replaced by $w_i$. This substitution ensures that the users outside of \arD, i.e, $w_i=0$, neither receive incentives, nor moves, which yields $\Pi_i^a= 0$. In other words, it turns on and off the triangle method, depending on the location of the user. Similarly, $u_i$ and $w_i$ are used to switch between~\eqref{c:coverage} and \eqref{c:coverage2}, based on the location of the user. 

The triangular method increases the number of variables by adding $3nm+1$ variables for each user, which makes it computationally costly for a large number of users. On the other hand, approximation quality of the triangle method increases with increasing number of breakpoints~\cite{dambrosio_piecewise_2010}. Therefore, an intervening SNC method, which maintains accuracy of JSNC and computational efficiency of USNC, is proposed next.

\section{Semi-Joint Spatial Network Configuration}\label{sec:Semi}
In Section~\ref{sec:USNC}, USNC is described, where the 3D placement of a drone-BS is performed without considering the effect of the users in \arD (Fig.~\ref{fig:range}). A brief analysis in Section~\ref{sec:single_user} showed that the maximum profit from a user decreases significantly as the distance between the user and drone-BS increases. Therefore, a JSNC method is developed in the preceding section, where the 3D placement is performed by taking the users outside the potential coverage region of drone-BS into account. However, JSNC yields a complex MINLP, which is computationally costly. Therefore in this section, we discuss a semi-JSNC method, which locates the drone-BS in 3D by considering the approximate effect of the users in \arD. Then, optimal incentive is offered to users based on~\eqref{eq:tau_star}. In other words, incentive design is not a completely inherent part of semi-JSNC, yet, it is straight forward to design incentives after carefully placing the drone-BSs.

Semi-JSNC aims at maximizing~\eqref{eq:piece} described in the preceding section with a slightly different approach on~\eqref{eq:inD}. The analysis in Section~\ref{sec:single_user} reveal that the maximum value of $\Pi(\tau_i, d_i)$ for a certain $d_i$ can be obtained by substituting $\tau_i^*$ corresponding to $d_i$. Taking~\eqref{eq:tau_star} into account, $\Pi(\tau_i^*, d_i)$ becomes a uni-variate function of $d_i$. Then, a piece-wise linear relationship can be obtained
\begin{numcases}{\sq{\check{\Pi}(d_i) = }\label{eq:uni_piece_wise}} 
p_1d_i + s_1, \ \ \ 0 < d_i \leq t_2, \\
p_2d_i + s_2, \ \ \ t_2 < d_i \leq t_3, \\
\qquad \  \vdots \quad \  ,  \qquad \quad \vdots \qquad \ \ \ ,\nonumber \\
p_Nd_i + s_N,  \ \, t_{N-1} <  d_i \leq d_u,\\ 
0, \  \qquad \qquad \ \text{o.w.}
\end{numcases}%
where $t_j$ for $j = 1,...,N$ denotes $N$ breakpoints on the $d-$axis, where $t_1 = 0 $, and $t_N = d_u$. The coefficients of the linear equation at each interval are denoted by $p_j$ and $s_j$. The selection and number of breakpoints determines the accuracy of~\eqref{eq:uni_piece_wise}. One of the common practices is choosing equally spaced breakpoints. However, adapting the length of the intervals to the characteristics of the function, e.g., taking breakpoints more frequently when the function changes rapidly, can provide better accuracy compared to the same number of equally-spaced breakpoints. 
In order to measure accuracy of approximations, RMSE is considered widely. An optimization problem to determine breakpoints with the objective of minimizing RMSE can be formulated by using~\eqref{eq:uni_piece_wise},
\begin{subequations}\label{eq:piece_optim}
\begin{align}
\tag{\ref{eq:piece_optim}}\label{eq:piece_optim_2}
&\underset{\parbox{5.5em}{\centering \scriptsize $\{p_j, s_j, t_j\}$}}{\text{minimize}}
\quad \int\left(\Pi(d_i) - \check{\Pi}(d_i; p_j, s_j, t_j)\right)^2 dt \\
& \quad \text{s.t.}
\qquad \qquad \ 0 \leq t_j \leq d_u, \qquad \forall j=2,...,N\!-\!1,\label{c:piece1}
\end{align}
\end{subequations}
where the first breakpoint, $t_1$, and the last breakpoint, $t_N$, are given (0 and $d_u$, respectively, in this case), and \eqref{c:piece1} ensures that the other breakpoints are within the first and last ones. In Fig.~\ref{fig:rmse}, piece-wise linear approximations with $N=3$ and $N=4$ breakpoints are shown with the corresponding RMSE values\footnote{Shape Language Modelling Toolbox for MATLAB is used to obtain the breakpoints. Available online: \textit{https://www.mathworks.com/matlabcentral/fileexchange/24443-slm-shape-language-modeling}, Accessed: 05 Sept., 2017.}.

A linear optimization problem can be designed for separable functions like~\eqref{eq:uni_piece_wise} by introducing continuous variables $\lambda_1, \lambda_2, ..., \lambda_N$ for each breakpoint, such that
\begin{align}
&\sum\limits_{j=1}^{N} \lambda_j\Pi(t_j) = \check{\Pi}(d_i) \label{c:equal_pi}\\
&\sum\limits_{j=1}^{N} \lambda_jt_j = d_i \label{c:equal_d}\\
&\sum\limits_{j=1}^{N} \lambda_j = 1\label{c:equal_1}
\end{align}
with the additional requirement that $\{\lambda_i\}_{i=1}^N$ are SOS2 type variables. It is required to guarantee that approximated values, $d_i$ and $\check{\Pi}(d_i)$ lie on the approximating line segments. 
Then, making use of the variables $u_i$ and $w_i$ indicating regions in Fig.~\ref{fig:range} (in Section~\ref{sec:JSNC}), an efficient optimization problem can be formulated 
\begin{subequations}\label{eq:fal_piece1}
\begin{align}
\tag{\ref{eq:fal_piece1}}\label{eq:obj_piece}
&\underset{\parbox{5.5em}{\centering \scriptsize $x_D,$ $y_D,$ $\{d_i,\check{\Pi}(d_i), u_i, w_i\}$ $\{\lambda_{i,j}, z_{i,j}\}$}}{\text{maximize}}
\qquad \sum_{i \in \mathbb{U}}u_i+\check{\Pi}(d_i)\\
& \text{s.t.}
\ \ \ \sq{r_i \leq \Gamma^* + M(1-u_i),}
\quad \qquad \qquad \, \forall i\! =\! 1,...,|\mathbb{U}|,\label{c:cov_in_R}\\
&\qquad \, \sq{r_i-d_i  \leq \Gamma^* + M(1-w_i),}
\quad \quad \ \ \ \ \forall i \! = \! 1,...,|\mathbb{U}|,\label{c:cov_in_D}\\
&\qquad \sum_{j=1}^{N} \lambda_{i,j}\Pi(t_j) = \check{\Pi}(d_i), \qquad \quad \ \ \ \forall \sq{i\! =\! 1,...,|\mathbb{U}|},\label{c:lam_pi}\\
&\qquad \sum_{j=1}^{N} \lambda_{i,j}t_j = d_i, \qquad\qquad\qquad\quad \, \forall \sq{i\! =\! 1,...,|\mathbb{U}|}, \label{c:lam_d}\\
&\qquad  \sum_{j=1}^{N}\lambda_{i,j} = w_i, \qquad\qquad\qquad\quad \ \ \forall i\! =\! 1,...,|\mathbb{U}|,\label{c:lam_w}\\
&\qquad \sum_{j=1}^{N} z_{i,j} \leq 2, \qquad\qquad\qquad\qquad \ \, \forall i\! =\! 1,...,|\mathbb{U}|,\label{c:sos1}\\
&\resizebox{0.86\columnwidth}{!}{
   $\displaystyle{\quad \  \sum_{k=j+2}^{N}\! z_{i,j}+z_{i,k} \leq 1, \quad  \forall j\! = \!1,...,N, \ \ \ \forall i\! =\! 1,...,|\mathbb{U}|,\label{c:sos2}}$}\\
&\qquad \lambda_{i,j} \leq z_{i,j}, \qquad\quad \ \ \forall j\!=\!1,...,N, \ \forall i\! =\! 1,...,|\mathbb{U}|,\label{c:sos3}\\
& \qquad  \lambda_{i,j} \in [0,\,1], \qquad\quad  \forall j\! =\! 1,...,N, \  \forall i\! =\! 1,...,|\mathbb{U}|,\\
&\resizebox{0.87\columnwidth}{!}{
   $\displaystyle{\quad \ \ z_{i,j}, u_i, w_i \in \{0,1\}, \  \forall j\! =\! 1,...,N, \  \forall i\! =\! 1,...,|\mathbb{U}|,\label{c:lam_son}}$}\\
& \qquad  x_{l} \leq x_D \leq x_{u},\\
& \qquad  y_{l} \leq y_D \leq y_{u},
\end{align}
\end{subequations}
where $z_{i,j}$ are binary variables used to ensure that $\lambda$'s are SOS2 variables via constraints~\eqref{c:sos1}, \eqref{c:sos2}, and \eqref{c:sos3}. The first two constraints are the same as in~\eqref{ob:upsi}, where $u_i$ and $w_i$ yield an objective function similar to~\eqref{eq:piece}. The piece-wise approximation constraints \eqref{c:lam_pi} and \eqref{c:lam_d} correspond to~\eqref{c:equal_pi} and \eqref{c:equal_d}, however, \eqref{c:equal_1} is modified to toggle the approximation via $w_i$ in \eqref{c:lam_w}. 

Semi-JSNC method reduces computational complexity by decreasing number of variables and constraints. One of the most significant changes from previous problem formulations is that $\tau_i$ is not an optimization variable any more. Instead, the incentive offered to a user is determined based on~\eqref{eq:tau_star}. Moreover, there are $N+2$ binary variables for each user, instead of having $nm\!+\!2$ of them in~\eqref{p:upsi}. Since $\check{\Pi}(d_i)$ is univariate, $N$ can be much less than $nm$.  Also, the number of equality constraints are reduced from 5 to 3 per user compared to JSNC. Run-time comparisons and numerical results of the presented SNC methods are discussed in the following section.  
\section{Simulations and Results}\label{sec:sim}
\begin{figure}[!t]
\centering
\includegraphics[width=0.45\textwidth, height=0.45\textwidth]{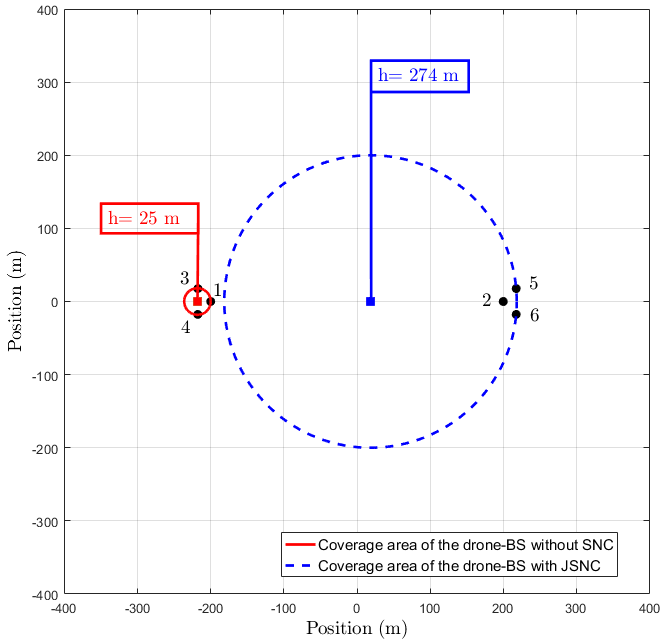}
\caption{A simple example demonstrating the difference of placement and profit with and without SNC framework. Normalized profit is 3 without SNC, and 4.16 with JSNC.}
\label{fig:toy}
\end{figure}
We consider the dense urban environment whose parameters are provided in Table~\ref{tab:param}. For the simulations, we assume one drone-BS is to be positioned. We also assume that the drone-BS utilizes orthogonal resources to any existing terrestrial infrastructure. Due to limited resources of drone-BSs, only users that cannot be served by the existing terrestrial networks are offloaded to the drone-BS. Additional simulation parameters are provided in Table~\ref{tab:sim}. 

\begin{figure}[!t]
\centering
\includegraphics[width=0.45\textwidth]{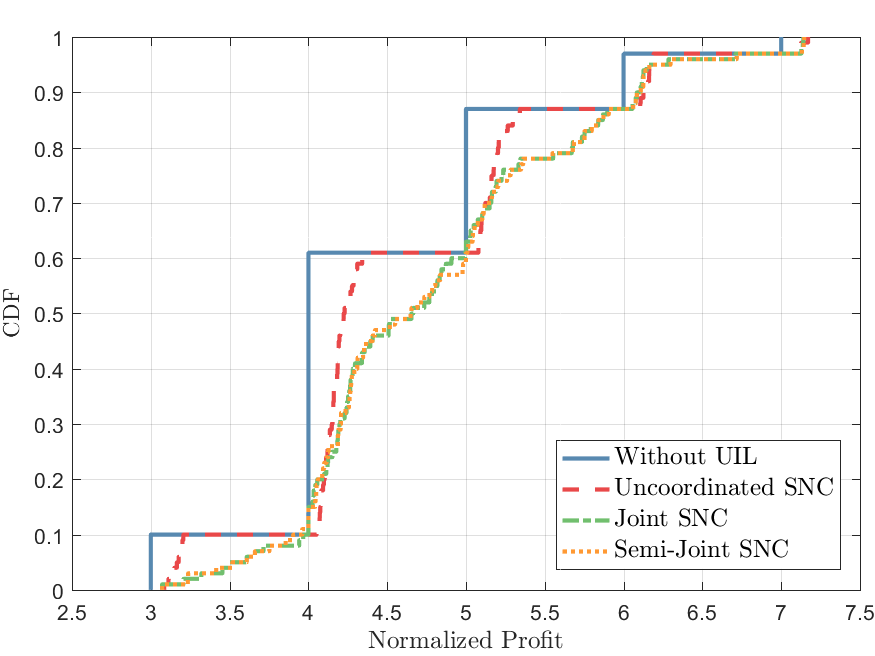}
\caption{CDF of the profit obtained by the proposed methods.}
\label{fig:profit_cdf}
\end{figure}

\begin{figure}[!h]
\includegraphics[width=0.48\textwidth]{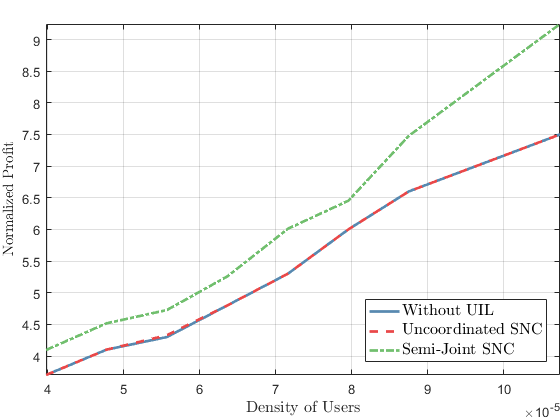}
\caption{The comparison of the proposed methods based on mean unit profit versus the density of the users.}
\label{fig:mean_profit}
\end{figure}

First, we would like to demonstrate the benefits of the proposed SNC framework utilizing drone-BSs and UIL by using a simple example. In Fig.~\ref{fig:toy}, users 1 and 2 are 400 m apart. Users 3\&4 and 5\&6 are 25 m apart from users 1 and 2, respectively. The provided environment and QoS parameters in Table~\ref{tab:sim} yields a maximum coverage radius of 200 m for a drone-BS with fixed transmit power. Therefore, the drone-BS cannot cover all the users while satisfying their QoS requirements. As seen in Fig.~\ref{fig:toy}, the efficient drone-BS placement method described in Section~\ref{sec:3d} covers only one of the groups with the minimum required coverage area (red circle with dotted line), which is the optimal solution in this scenario without SNC. It provides a normalized profit of 3. On the other hand, the proposed JSNC method covers  the group of users 2, 5, and 6, and offers incentives to the remaining users (blue circle with dotted line). Note that, not only coverage radius is expanded to 200 m, but also altitude, and horizontal location of the drone-BS is adjusted. Hence, JSNC provides total normalized profit of 4.16, which is $39\%$ more than the previous method. 
\begin{table}[!t]
\caption{Simulation Parameters}
\label{tab:sim}
\centering
\begin{tabular}{|c|c|}
\hline
\textbf{Parameter} & \textbf{Value}\\
\hline
($x_{l}, x_{u}$) & (-700, 700) m\\
\hline
($y_{l}, y_{u}$) & (-700, 700) m\\
\hline
$\gamma$ & 90 dB\\
\hline
$f_c$ & 2.5 GHz\\
\hline
$d_{u}$ & 200 m\\
\hline
$\mathrm{W}$ & 700 m \\
\hline
$\Gamma^*$ & 209 m \\
\hline
Monte Carlo Runs & 100\\
\hline 
$\tau$ vertices, $\tilde{\tau}$ & $\{0.05,\, 0.1,\, 0.2,\, 0.9\}$\\
\hline
$d$ vertices, $\tilde{d}$ & $\{5,\, 10,\, 20,\, 40,\, 200\}$\\
\hline
Number of breakpoints for Semi-JSNC, $N$ & 3\\ 
\hline
Number of users, $N_u$ & 15\\
\hline
\end{tabular}
\end{table}
\begin{figure}[!t]
\centering
\includegraphics[width=0.45\textwidth]{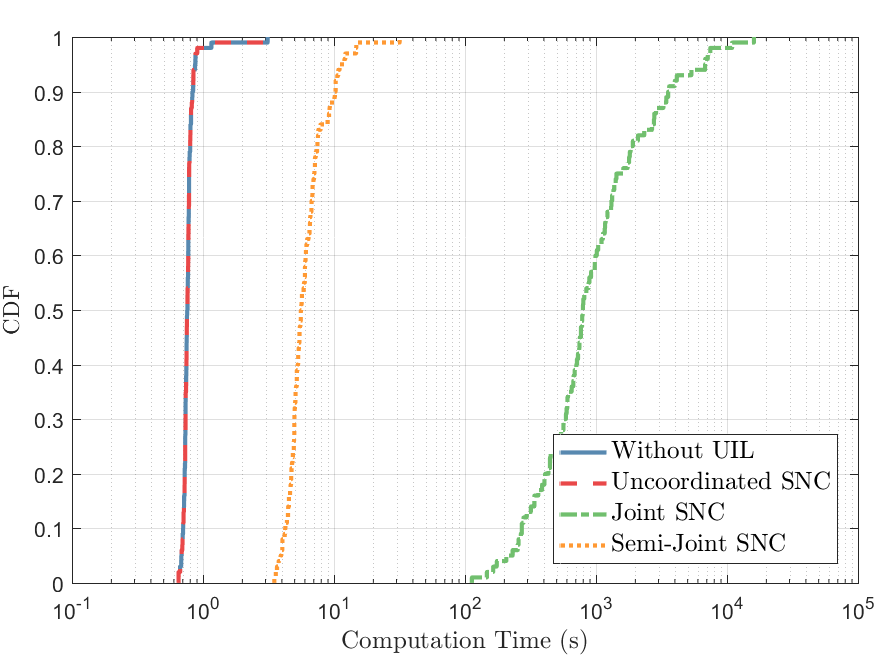}
\caption{CDF of the computation time of the proposed methods.}
\label{fig:time}
\end{figure}
Although the SNC can be beneficial for future wireless networks, it can increase computational complexity for 3D drone-BS placement. Note that it is possible to have the placement and incentive design in a centralized entity, and configure the drone-BSs accordingly. One can consider this system as the centralized \textit{drone management framework} proposed in~\cite{bor_magazine}. Therefore, the complexity of the algorithm may not have any affect for the drone-BS itself. Nevertheless, computationally efficient algorithms are more favorable for the overall computational burden of wireless networks. In order to address this issue, three methods are proposed in this article with varying computational complexities. The performance of these proposed methods is compared by randomly placing 15 users in $\mathcal{W}$ (Fig.~\ref{fig:range}) with parameters provided in Table~\ref{tab:sim}. The cumulative distribution function (cdf) of the computation time of each method is given in Fig.~\ref{fig:time}. Note that, the computational complexity of the USNC is similar to that of the scheme without UIL, because once the drone-BS is placed, the optimal incentive to the users within the extended coverage area ($\mathcal{D}$ in Fig.~\ref{fig:range}) is calculated simply by using (16). On the contrary, JSNC method, which considers the placement of the drone-BS jointly with the incentive amount to be offered to each user, has the highest computational complexity. Since the JSNC is a computationally costly method, semi-JSNC is proposed in Sec. V. As shown in Fig.~\ref{fig:time}, semi-JSNC reduces computational time by more than 2 orders of magnitude. Success of semi-JSNC is mainly due to reduction of variables and constraints, as discussed in the end of Section IV and V. Table~\ref{tab:vars} shows number of simulation parameters for 15 users. Note that cones correspond to quadratic constraints, which is the same in both JSNC and semi-JSNC, due to addition of the second quadratic constraint for distance of users outside the coverage area of the drone-BS. On the other hand, number of all other parameters is significantly reduced when comparing JSNC and semi-JSNC. 
At the same time, the performance of the semi-JSNC method is very close to that of the JSNC as Fig.~\ref{fig:profit_cdf} demonstrates. Both JSNC and semi-JSNC outperform USNC in terms of normalized profit.
\begin{table}[t!]
\centering
\caption{Number of simulation parameters for 15 users}
\begin{tabular}{l|l|l|l|}
\cline{2-4}
                       & USNC  & JSNC  &  Semi-JSNC  \\ \hline
\multicolumn{1}{|l|}{ Constraints} &  43  &  1108  &  254  \\ \hline
\multicolumn{1}{|l|}{ Cones} &  15  &  30 &  30  \\ \hline
\multicolumn{1}{|l|}{ Scalar variables} &  90  &  1650  &  390  \\ \hline
\multicolumn{1}{|l|}{ Integer variables} &  15  &  780 &  75  \\ \hline
\end{tabular}\label{tab:vars}
\end{table}
Finally, Fig.~\ref{fig:mean_profit} shows the  mean normalized profit with respect to the user density. The user density is calculated as
\begin{equation}
\frac{N_u}{\pi \mathrm{W}^2}.
\end{equation} 
In this experiment, $ \mathrm{W} = 700$ m and a varying number of users from 10 to 27 are distributed randomly. For each number of users, 10 simulations are conducted, and the average of these simulations is calculated. Fig.~\ref{fig:mean_profit} shows that semi-JSNC outperform USNC and the scheme without UIL, where the latter two provide very similar results. This indicates that USNC is not enough to attract enough number of users to provide a significant gain, and the situation does not improve with increasing density of users. On the other hand, gain from semi-JSNC increases with increasing user density, relative to the USNC gain. Note that, JSNC is not depicted in this figure, due to computational complexity. 
\\
\section{Conclusions}\label{sec:conc}
In this paper, we proposed a novel method, SNC, based on exploiting the mobility of both the users and the drone-BSs. We first propose an uncoordinated SNC method, which is based on placing the drone-BS in 3D space, and then offering the incentives to users. Although USNC is computationally efficient, the performance gain from considering these elements (drone-BS and UIL) separately provides a limited gain in terms of number of served users (added capacity to the system), because the drone-BS is not positioned by considering potential gain from UIL. In order to alleviate this issue, we proposed a joint-SNC method, which can be used to obtain the optimal incentives and 3D coordinates simultaneously. However, the gain from UIL increases at the cost of computational complexity. Therefore, we propose a semi-joint SNC method, which is computationally more efficient compared to joint-SNC, and provides similar results. 

Once drone-BSs are efficiently utilized via SNC, their flexibility and customizable nature can make them suitable to enable agile networking, which is useful for many futuristic scenarios such as IoT, tactile internet, or green networking. Especially the area of green networking is worth investigating thoroughly, since energy-efficiency oriented SNC schemes can help alleviate for the excess energy expenditure of drone-BSs. However, more comprehensive inter-disciplinary studies are needed to improve understanding of wireless network users' preferences for cooperating with network operators. Moreover, the placement and incentive options presented here are not necessarily ultimate placements and incentives, rather, they can serve as good initializations for adaptive algorithms that we are hoping to explore in the future. Finally, since the proposed SNC schemes are versatile enough to apply in various settings, investigating trade-offs for multiple-drone-BSs and placement methods for area coverage with SNC can be listed among other interesting future research directions.

\begin{appendices}
\section{Explanation of Observation 1}\label{app:A}
\begin{figure}[!t]
\centering
\includegraphics[width=0.45\textwidth]{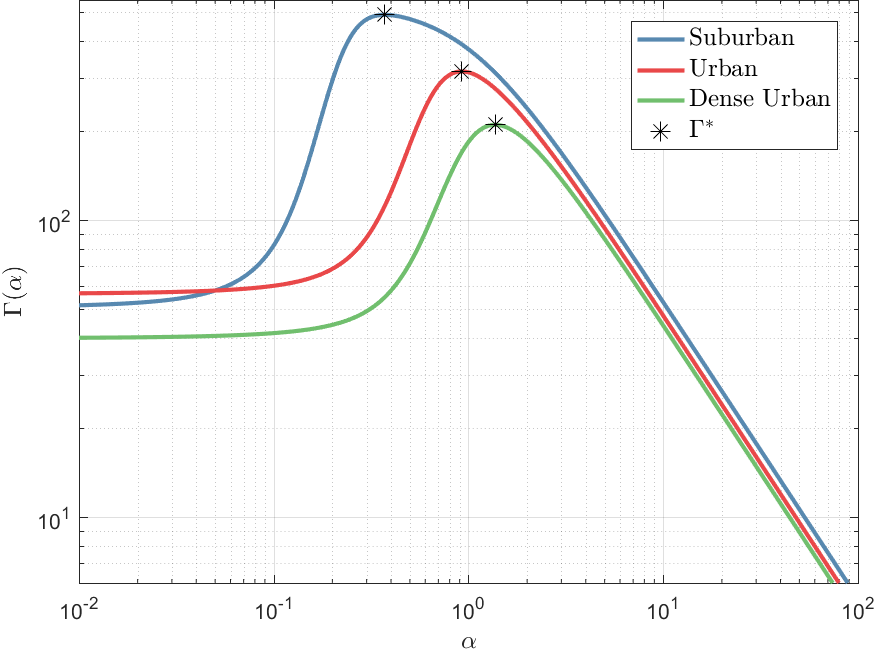}
\caption{$\Gamma(\alpha)$ versus $\alpha$ for the environments in Table~\ref{tab:param}.}
\label{fig:All_env}
\end{figure}

The largest feasible set can be obtained for the largest value of $\Gamma(\alpha)$, which is $\Gamma^*$. Note that, the behaviour of $\Gamma(\alpha)$ is only determined by the environment parameters $a$ and $b$, and $\alpha$, whereas the other parameters simply scale the function. $\Gamma^*$ can be found by solving $\frac{d\Gamma(\alpha)}{d\alpha} = 0$, which can be written as
\begin{equation}
\label{eq:gammader}
\resizebox{0.90\columnwidth}{!}{$\displaystyle{\dfrac{k_3{\cdot}10^\frac{\frac{\!-\!z_1}{P(\alpha)}-z_2+\gamma}{20}\sq{\mathrm{e}^{\!-b\left(\frac{180\tan^{-\!1}\left(\alpha\right)}{{\pi}}-a\right)}}}{{\pi}\!\left(\!\alpha^2\!+\!1\right)^\frac{3}{2}\!\left(\!a\mathrm{e}^{-b\left(\frac{180\tan^{-\!1}\left(\alpha\right)}{{\pi}}-a\right)}\!+\!1\!\right)\!^2}-
\dfrac{\alpha{\cdot}10^\frac{\frac{\!-z_1}{P(\!\alpha\!)}\!-\!z_2\!+\!\gamma}{20}}{\left(\alpha^2+1\right)^\frac{3}{2}} = 0 \numberthis}$}.
\end{equation}
After manipulations, the $\alpha*$ can be obtained as~\eqref{eq:alpha_star}.
Unfortunately, it is not possible to obtain a closed for expression for $\alpha^*$. By numerically plotting~\eqref{eq:Gamma} in  Fig.~\ref{fig:All_env}, we show that for all environments there exists only one maximum value, which occurs at $\alpha^*$. Moreover, it is observed that the local maximas marked in Fig.~\ref{fig:All_env} are the only maximas for all environments. Note that, the considered interval for $\alpha \in [0.01, 100]$ in Fig.~\ref{fig:All_env} includes the practical values for low and medium altitude platforms~\cite{handbook}.

♣

\section{Regional Analysis on Coverage and Optimal Incentive}\label{app:B}

Coverage of a drone-BS is investigated in several studies, e.g., \cite{hayajneh2016drone} and \cite{al-hourani_optimal_2014}, however, UIL is not considered in any of the studies so far. As discussed in Section~\ref{sec:uil}, there is a probability that the user will move from \arD ~to \arR, which means that the previously uncovered users can be covered under certain circumstances. 

Regional analysis can be useful when the location information of users is not available, hard to obtain, or incorrect, when the users are highly mobile such that the coherence time of snapshot analysis is very low, or when the profitability of the region is to be investigated for a known user density.

\subsection{Extended Coverage of a BS in a System with UIL}
Let $\text{P}^\text{R}$ denote probability of coverage for a user $i$ in \arR. Since users are uniformly distributed, and a user is assumed to be covered if it is in \arR,
\begin{align*}
\label{eq:pR}
\text{P}^\text{R} &= \text{P}\lbrace u_i\! =\! 1 \,\cap\, (x_i,y_i)\! \in\! \mathcal{R}\rbrace\\
& = \frac{R^2}{W^2}, \numberthis
\end{align*}
where $u_i = 1$ indicates that the user $i$ is covered, and $u_i=0$ indicates the opposite. 

The users within range $D$ can also be covered, if they accept the incentive and move in \arR. Let $\text{P}^\text{D}$ denote the probability of a user $i$ to be covered in \arD,
\begin{align*}
\label{eq:pD}
\text{P}^\text{D} &= \text{P}\lbrace u_i\! =\! 1 \,\cap\, (x_i,y_i)\! \in\! \mathcal{D}\rbrace\\
& = \int_{R}^{R+d_u} \frac{2\pi r}{\pi W^2} \mathrm{e}^{-\beta(\tau)(r-R)} dr\numberthis\\
& = \resizebox{0.58\columnwidth}{!}{$\displaystyle{\sq{ \frac{2\left((-\beta(\tau)(R+d_u)-1)\mathrm{e}^{-\beta(\tau)(d_u)}+\beta(\tau)R+\!1\right)}{W^2\beta ^2 (\tau)}}}$}
\end{align*}%
Then, coverage probability of any user in \arW, i.e., $(x_i,y_i) \in \mathcal{W}$, can be obtained from \eqref{eq:pR} and \eqref{eq:pD} as
\begin{equation}
\label{eq:prob_cov}
\text{P}^\text{W} = \text{P}^\text{R} + \text{P}^\text{D}. \numberthis\\
\end{equation}
The coverage probabilities calculated above, help analysing profitability of a region, as discussed in the following section.

\subsection{Optimal Regional Incentive}
\label{sec:opt_incent}
\begin{figure}[!t]
\centering
\includegraphics[width=0.48\textwidth]{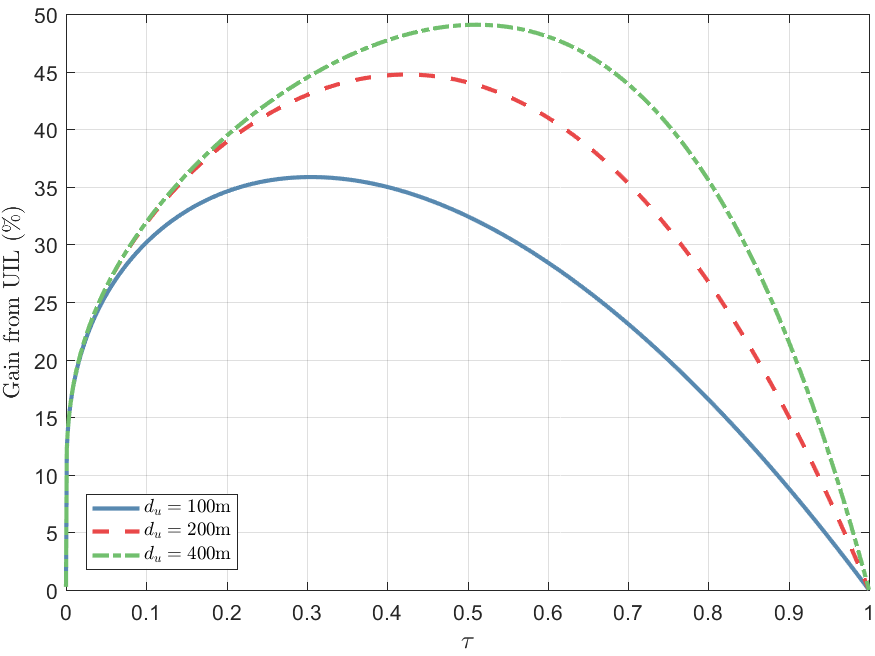}
\caption{Gain from the UIL with respect to regional incentive, $\tau$ is calculated for $R\,=\,200$, $W\,=\,2000$.} 
\label{fig:profit}
\end{figure}

The profit of the network operator is the difference between the income from all the covered users, and the cost of the incentives. Assuming all users in \arD~receive the same incentive, which is the \textit{regional incentive}, $\tau$, the average normalized revenue per user is
\begin{align}
\label{eq:pi_not}
\Pi_{\circ}(\tau) &= \text{P}^\text{R} + \text{P}^\text{D}(1-\tau). \numberthis
\end{align}
The percentage of the gain from UIL, which is $100\times\frac{\text{P}^\text{D}(1-\tau)}{\text{P}^\text{R}}$, is plotted in Fig.~\ref{fig:profit}. As can be observed from Fig.~\ref{fig:profit}, an optimal regional incentive maximizing the profit, $\tau^*$, can be obtained. The main reason for this behaviour is that the probability of moving for a used (given in~\eqref{eq:prob_move}) does not increase significantly even if more incentive is offered. The optimal regional incentive $\tau^*$ can be obtained by solving $\frac{d\Pi_{\circ}(\tau)}{d\tau} = 0$, which is the solution of
%
\begin{align}
&\resizebox{0.97\columnwidth}{!}{$\displaystyle{2\beta(\!\tau\!)k_1\!\left(\!-d_u\mathrm{e}^{\!-\!\beta(\!\tau\!)d_u}\!\left(\!(R+d_u)(\!-\beta(\tau)\!-\!1)\right)\!-\!(R+d_u)\mathrm{e}^{\!-\!\beta(\!\tau\!)d_u}\!+\!R\right)-}$} \nonumber\\
&\resizebox{0.82\columnwidth}{!}{$\displaystyle{4k_1\left( \mathrm{e}^{\!-\!\beta(\!\tau\!)d_u}\left(\!(R+d_u)(\!-\beta(\tau)-1)\right) + R\beta(\tau)\!+\!1\!\right) = 0}$}. 
\end{align}
%

\begin{figure}[!t]
\centering
\includegraphics[width=0.48\textwidth]{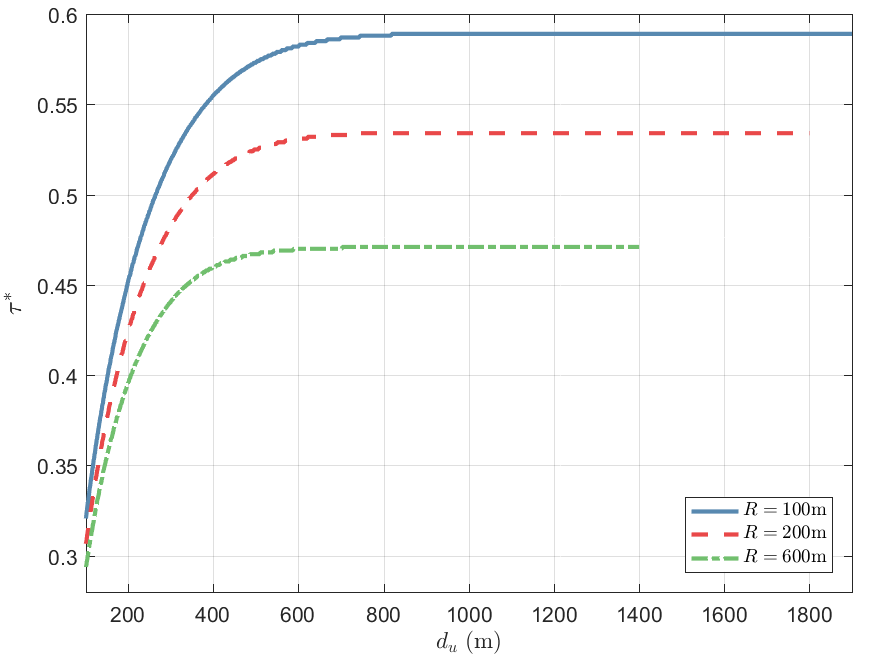}
\caption{Optimal regional incentive is calculated with respect to $d_u$ for $R\,=\,\{ 100,\, 300,\, 500,\, 700\} $, $W\,=\,2000$.} 
\label{fig:x_d}
\end{figure}

Again from Fig.~\ref{fig:profit}, we observe that both the gain from the UIL and $\tau^*$ increases with increasing $d_u$ for a given coverage radius, $R$. Therefore, in Fig.~\ref{fig:x_d}, we consider $d_u \rightarrow W-R$ for a given $R$. It means that $d_u$, and, hence, \arD~is as large as possible, which should lead to maximum gain from UIL. We observe that $\tau^*$ saturates to a certain value, $\tau^\infty$, which will be obtained by analysing limiting conditions of~\eqref{eq:pi_not} next. 

First, note that the average normalized revenue in a region with $N$ users is 
\begin{equation}
\Pi_N(\tau) = N\times\Pi_\circ(\tau).
\end{equation}
In order to obtain an infinite region, consider $W \rightarrow \infty$ and $d_u \rightarrow W-R $. Then, the limit of $\text{P}^\text{R} \rightarrow 0$, as can be observed from ~\eqref{eq:pR}. Let $\Pi_N^\infty(\tau)$ denote the limit of $\Pi_N(\tau)$ such that
\begin{equation}
\lim_{\substack{W \rightarrow \infty, \\ d_u \rightarrow W-R }} \Pi_N(\tau) = \Pi_N^\infty(\tau). 
\end{equation}
For a fixed user density, $\lambda$, substituting $N = \lambda \pi W^2$, results
\begin{equation}
\Pi_N^\infty(\tau)=\lambda \pi \left( R^2\! +\! \frac{2\left(1\!-\!\tau\right)}{\beta^2(\tau)}\left(\!-\!\beta(\tau)R+1\right)\right).
\label{eq:pi_inf}
\end{equation}
Then, the optimal regional incentive maximizing $\Pi_N^\infty(\tau)$, $\tau^\infty$, is simply the solution of $\frac{d\Pi^{\infty}(\tau)}{d\tau} = 0$, which is 
\begin{align*}
\sq{
\frac{k_1 R\! \left( 1\!-\!\tau\right)}{\tau\beta^2(\tau)} 
- \frac{R\beta(\tau)+1}{\beta^2(\tau)}
+ \frac{2k_1 \left(1\!-\! \tau\right)\left(\!-\!R\beta(\tau)+1\right)}{\tau\beta^3(\tau)} = 0\numberthis.  }
\label{eq:x_opt}
\end{align*}
Note that, $\Pi_N^\infty(\tau^\infty)$ represents the maximum gain from UIL for any given coverage radius, $R$, and user density, $\lambda$. Moreover, $\Pi_N^\infty(\tau^\infty)$ is valid for any base station, either terrestrial or aerial. In case of drone-BSs,  $\Pi_N^\infty(\tau^\infty)$ can be used to analyse the cost of expanding the coverage region, which can either be performed by increasing the transmit power, which may lead to changing the type of the utilized drone, required legal permissions and so on~\cite{bor_magazine}, or simply offering more incentives to the users.

%
%
%

%


\label{app:main}
\end{appendices}

%


\bibliographystyle{IEEEtran}
\bibliography{droneUIL}

%

%
%
%




\end{document}